\documentclass[aps,floats,twocolumn,prd,nofootinbib,superscriptaddress]{revtex4-1}
\usepackage[utf8]{inputenc}
\usepackage{bm}
\usepackage{amsmath}
\usepackage{comment}
\usepackage{color}
\usepackage{natbib}
\usepackage{float}
\usepackage{graphicx}

\def\VEV#1{{\left\langle #1 \right\rangle}}

\begin{document}

\title{Pulsar-timing arrays, astrometry, and gravitational
waves}

\author{Wenzer Qin}
\author{Kimberly K. Boddy}
\author{Marc Kamionkowski}
\affiliation{Department of Physics and Astronomy, Johns Hopkins
     University, 3400 N.\ Charles St., Baltimore, MD 21218, USA}
\author{Liang Dai}     
\affiliation{School of Natural Sciences, Institute for
     Advanced Study, Princeton, NJ 08540}

\begin{abstract}
We discuss the theory of pulsar-timing and astrometry probes of a
stochastic gravitational-wave background with a recently
developed ``total-angular-momentum'' (TAM) formalism for cosmological
perturbations.  We review the formalism, emphasizing in
particular the features relevant for this work and
describe the observables we consider (i.e.
the pulsar redshift and stellar angular displacement).  Using the
TAM approach, we calculate the angular power spectra for the
observables and from them derive angular auto- and
cross-correlation functions.  We
provide the full set of power spectra and correlation functions
not only for the standard transverse-traceless propagating
degrees of freedom in general relativity, but
also for the four additional non-Einsteinian polarizations that
may arise in alternative-gravity theories.  We discuss how 
pulsar-timing and astrometry surveys can complement and serve as
cross checks to one another and comment on the
importance of testing the chirality of the gravitational-wave
background as a tool to understand the nature of its sources.
A simple rederivation of the power spectra from the plane-wave
formalism is provided in an Appendix.
\end{abstract}

\maketitle

\section{Introduction}

Efforts to detect a stochastic gravitational-wave background
using pulsar-timing arrays have been around for almost three
decades \cite{Foster:1990}.  There are now three major
efforts: the Parkes Pulsar Timing Array (PPTA)
\cite{Hobbs:2013aka,Manchester:2012za}, North American Nanohertz
Observatory for Gravitational Waves (NANOGrav)
\cite{Arzoumanian:2018saf}, and the European Pulsar Timing Array
(EPTA) \cite{Lentati:2015qwp}.  The three collaborate through an
International Pulsar Timing Array (IPTA)
\cite{Verbiest:2016vem}.  The effects of gravitational waves on
the pulse arrival times from pulsars were worked out presciently
first in Refs.~\cite{Detweiler:1979wn,Sazhin:1978}.  The
signature of a stochastic gravitational-wave background is the
characteristic angular correlation in the timing residuals
worked out by Hellings and Downs \cite{Hellings:1983fr}.  The
measurements, which span timescales of years, are sensitive
primarily to gravitational waves with frequencies
$\sim10^{-9}$~sec$^{-1}$, though constraints at lower frequencies
have been considered as well~\cite{Zakamska:2005hn}.  The endeavor is particularly exciting
given that a stochastic background in this frequency range is
expected from the mergers of supermassive-black-hole binaries
\cite{Rajagopal:1994zj,Jaffe:2002rt,Sesana:2012ak,Ravi:2014nua,Kelley:2017lek}.
The first data release from IPTA has placed a $2\sigma$ limit
on the dimensionless strain of the stochastic background to be
$1.7 \times 10^{-15}$ at a frequency of 1~yr$^{-1}$, with an
assumed spectral index of $-2/3$, and significant improvement
in sensitivity is expected with the next dataset \cite{Verbiest:2016vem}.
See Refs.~\cite{Maggiore:1999vm,Burke-Spolaor:2015xpf,Lommen:2015gbz,Hobbs:2017oam,Yunes:2013dva}
for recent reviews of the effort to 
detect gravitational waves with pulsar timing.

Attention has recently turned to the
possibility to detect a stochastic gravitational-wave
background with astrometry \cite{Braginsky:1989pv,Kaiser:1996wk},
which probes frequencies $H_0 \lesssim f \lesssim 1~\mathrm{yr}^{-1}$
($10^{-18}~\mathrm{s}^{-1} \lesssim f \lesssim 10^{-8}~\mathrm{s}^{-1}$)
that overlap with and bridge the frequency gap between
cosmic microwave background polarization measurements and
pulsar-timing measurements \cite{Darling:2018hmc}.  Book and Flanagan
\cite{Book:2010pf} provided the first detailed characterization
of the expected signals in terms of angular correlation
functions and power spectra.  Their work has been extended to
the search for point sources of gravitational waves
\cite{Moore:2017ity} and to non-Einsteinian polarizations
\cite{Mihaylov:2018uqm,OBeirne:2018slh}, the latter of which
echoes analogous work for pulsar timing \cite{Lee:2008aa,Chamberlin:2011ev}.
Astrometric data from GAIA and extragalactic radio sources
constrain the energy density (integrated over $\ln f$) of the
stochastic background to be $< 0.011$ for frequencies
$6\times 10^{-18}~\mathrm{s}^{-1} \lesssim f \lesssim 10^{-9}~\mathrm{s}^{-1}$
\cite{Darling:2018hmc}.  Future
astrometry missions \cite{Boehm:2017wie} might provide improved
data for such measurements.

Here we extend previous work on the calculation of angular
correlation functions and angular power spectra for
pulsar-timing and astrometry probes of the gravitational-wave
background by employing a ``total-angular-momentum'' (TAM)
formalism developed recently \cite{Dai:2012bc,Dai:2012ma} for the study
of cosmological perturbations.  In most discussions of
cosmological perturbations and stochastic gravitational-wave
backgrounds, the spacetime-metric perturbation is decomposed
into plane waves $e^{i \boldsymbol{k}\cdot \boldsymbol{x}}$, as these provide a
simple and familiar complete orthonormal basis.  
For gravitational
waves in general relativity, there are two polarizations,
typically taken to be $+$ and $\times$, with polarization vectors
$\epsilon_{ab}^+(\boldsymbol{k})$ and
$\epsilon_{ab}^\times(\boldsymbol{k})$ associated with each
wave vector $\boldsymbol{k}$.  
The simplicity is lost, though, when
projecting these plane waves onto observables on the spherical
sky.  The TAM approach has been applied to simplify calculations
of weak gravitational lensing \cite{Dai:2012bc}, angular
three-point functions \cite{Dai:2012ma}, and circular
polarization of the cosmic microwave background
\cite{Kamionkowski:2018syl}.

As elaborated below, the TAM formalism provides an alternative
complete orthonormal set of basis functions: the TAM waves.  In
this formalism, the wave vector $\boldsymbol{k}$ is replaced by quantum
numbers $k\ell m$, where $k$ is a wave number magnitude (equivalent
to $\boldsymbol{k}$) and $\ell m$ are total-angular-momentum quantum
numbers.  Observables on the sphere are similarly quantified in
terms of quantum numbers $\ell m$, and any such observable receives
contributions only from TAM waves of the same $\ell m$.  This leads,
as we will see, to simple derivations of the predictions for
harmonic-space observables.  The $+$ and
$\times$ polarizations in the plane-wave expansion are replaced
in TAM waves by two transverse-tensor polarizations which we
call ``tensor E'' (TE) and ``tensor B'' (TB).  We decompose the
two scalar polarizations that 
may arise in alternative-gravity theories  into
scalar-transverse (ST), sometimes referred to as ``breathing'',
and scalar-longitudinal (SL) modes to correspond to the
decomposition used in prior work.  There are also two  vector
polarizations that we call ``vector E'' (VE) and ``vector B''
(VB).

Our paper is organized as follows:  in Sec.
\ref{sec:observables} we first describe our characterization of
the observables.  For pulsar timing, this is pulse frequency, and for
astrometry the angular positions, at two different epochs.  These
are then translated to spherical-harmonic coefficients from
which correlation functions and power spectra are derived.  We
provide the complete set of six two-point
angular correlation (and cross-correlation) functions
for a combined pulsar-timing/astrometry survey and relate them
to the six angular power spectra.  We summarize briefly our
main results in Sec. \ref{sec:summary} before going through
the calculations in Sec. \ref{sec:calcs}, first for the
redshift and then for astrometry.  This section also
discusses the results for the various power spectra and
autocorrelation functions.  Section
\ref{sec:crosscorrelation} presents results for the
redshift-astrometry cross-correlations.  In Sec.
\ref{sec:windowfns} we discuss the range of
gravitational-wave frequencies probed by pulsar timing and
astrometry, point out that information on the local
three-dimensional metric perturbation can be reconstructed from
combined angular/time-sequence information, and 
emphasize the importance of pursuing the parity-violating
observables that may arise from chirality in the
gravitational-wave background.  Section
\ref{sec:conclusions} provides concluding remarks.
In Appendix \ref{sec:TAM} we provide a brief reprise of
Ref.~\cite{Dai:2012bc}, emphasizing in particular the aspects
relevant for the study of a stochastic gravitational-wave
background, as well as a few new results needed for our
calculations. Appendix \ref{sec:legendretricks}
provides some Legendre-polynomial relations needed to translate
angular power spectra and angular correlation functions.
Appendix \ref{sec:HDsimple} describes a simple alternative
technique, based on the plane-wave formalism, to derive all of
the power-spectrum results.  Appendix
\ref{sec:angularcorrelations} derives the relations between
angular power spectra and correlation functions.

\section{Observables}
\label{sec:observables}

We begin by describing the observables.  For simplicity/clarity,
we assume that there are PTA and astrometry measurements performed at
two times $t$ and $t +\Delta t$ separated by a time interval
$\Delta t$.  The generalization to more realistic observational
cadences is described briefly later.

\subsection{Spherical-harmonic coefficients and power spectra}

We assume a multitude of pulsars spread throughout
the sky and that a pulsar in a direction $\boldsymbol{\hat{n}}$
is observed to
have a redshift $z(\boldsymbol{\hat{n}},t)$ at time $t$.  Since,
in practice, a single pulse is typically buried in noise and is
thus undetectable, the relevant observable is the timing residual
$\int^t dt'\, z(\boldsymbol{\hat{n}},t')$, obtained by
accumulating many pulses. To simplify the discussion in this paper,
we consider the observable to be the change $(\delta
z)(\boldsymbol{\hat{n}},t) \equiv z(\boldsymbol{\hat{n}},t+\Delta
t)-z(\boldsymbol{\hat{n}},t)$ over the time interval $\Delta t$.%
\footnote{Using the simplified observable $(\delta z)(\boldsymbol{\hat{n}},t)$
does not affect our main results, which relate to the angular dependence
of correlation functions.}
These observational ``data'' can be represented alternatively
and equivalently in terms of the spherical-harmonic coefficients,
\begin{equation}
     z_{\ell m}(t) = \int \,d\boldsymbol{\hat{n}} \, Y^*_{\ell m}(\boldsymbol{\hat{n}}) (\delta
     z)(\boldsymbol{\hat{n}},t),
\end{equation}
where $Y_{\ell m}(\boldsymbol{\hat{n}})$ are spherical harmonics.
If the $z_{\ell m}(t)$ are provided, the change in redshift
can be obtained from the inverse transformation,
\begin{equation}
     (\delta z)(\boldsymbol{\hat{n}},t) = \sum_{\ell=0}^\infty
     \sum_{m=-\ell}^\ell z_{\ell m}(t) Y_{\ell m}(\boldsymbol{\hat{n}}).
\end{equation}

For astrometry measurements, we assume that each source, with its
proper motion already accounted for, in the survey has moved an
angular distance $(\delta n)_a$ over the time interval $\Delta t$
due to the presence of a gravitational-wave background.
From such measurements for sources spread over the sky, we obtain a
deflection-angle field $(\delta n)_a(\boldsymbol{\hat{n}})$, which
is a vector field (represented with the
single abstract index $a$) that lives in the
celestial sphere and is a function of position on the sky.  It
can thus be expanded in vector spherical harmonics [as defined
in Eq.~(38) in Ref.~\cite{Dai:2012bc}],\footnote{Appendix B in
Ref.~\cite{Gair:2015hra} provides useful properties of these
harmonics, although their vector harmonics are smaller than ours
by a factor of $\sqrt{2}$.}
\begin{equation}
     (\delta n)_a(\boldsymbol{\hat{n}}) = \sum_{lm} \left[ E_{\ell m}
     Y^E_{(\ell m)a}(\boldsymbol{\hat{n}}) + B_{\ell m} Y^B_{(\ell m)a}(\boldsymbol{\hat{n}}) \right],
\end{equation}
in terms of spherical-harmonic coefficients,
\begin{eqnarray}
     E_{\ell m} &=& \int\, d\boldsymbol{\hat{n}}\, (\delta
     n)^a(\boldsymbol{\hat{n}}) 
     Y^E_{(\ell m)a}(\boldsymbol{\hat{n}}), \nonumber \\
     B_{\ell m} &=& \int\, d\boldsymbol{\hat{n}}\,
     (\delta n)^a(\boldsymbol{\hat{n}}) Y^B_{(\ell m)a}(\boldsymbol{\hat{n}}).
\end{eqnarray}

The values that the $z_{\ell m}$, $E_{\ell m}$, and $B_{\ell m}$ take depend
on how our coordinate system is chosen.  The power spectra
\begin{equation}
     C_\ell^{XX} = \frac{1}{2\ell+1} \sum_m \left| X_{\ell m} \right|^2,
\end{equation}
for $X=\{z,E,B\}$ are, on the other hand, rotational invariants.
Here $C_\ell^{EE}$ and $C_\ell^{BB}$ are
power spectra for, respectively, the E and B
modes.  There are three additional cross-correlation power spectra,
\begin{equation}
     C_\ell^{XX'} = \frac{1}{2\ell+1} \sum_m X_{\ell m}
     \left(X_{\ell m}^\prime \right)^*,
\end{equation}
for $XX'=\{zE,zB,EB\}$ that are also rotationally invariant.
The cross-spectrum $C_\ell^{EB}$ is expected, given
the opposite parities of E and B, to be zero unless the
gravitational-wave background breaks parity.
The redshift $z(\hat{n},t)$ is associated with the longitudinal
vector harmonic, which has the same parity as E.
Thus, we also expect $C_\ell^{zB}$ to be zero and $C_\ell^{zE}$
to be nonzero if parity is not broken.

If the signal is due to a statistically isotropic stochastic
background, then we expect
\begin{equation}
     \VEV{X_{\ell m} \left(X'_{\ell'm'} \right)^*} = C_\ell^{XX'}
     \delta_{\ell\ell'} \delta_{mm'},
\end{equation}
for all six $XX'=\{zz,EE,BB,zE,zB,EB\}$.  This expression says
that the variance of any $X_{\ell m}$ is $C_\ell^{XX}$, and the
covariance of any two different ones is $C_\ell^{XX'}$.  It also
tells us that each spherical-harmonic coefficient is
statistically independent.  If the background is moreover a
Gaussian random field (e.g., as arises for inflationary
gravitational waves), then each $X_{\ell m}$ (actually, its real and
imaginary components) is chosen from a Gaussian distribution.
In this case, the variance with which the theoretical
expectation for $C_\ell^{XX}$ can be obtained in the 
ideal case is $\sqrt{2/(2\ell+1)} C_\ell^{XX}$ (with analogous
expressions for the covariances as given, for example, in
Ref.~\cite{Kamionkowski:1996ks}).  In practice, the most 
likely background, from supermassive black holes (SMBHs), is unlikely to be Gaussian, and
so this cosmic variance will be a bit different~\cite{Taylor:2015udp,Mingarelli:2017fbe}.

The statistical independence of the $X_{\ell m}$ gives the harmonic
approach (i.e., working with the $X_{\ell m}$ and $C_\ell^{XX'}$) a
conceptual advantage over the configuration-space approach
[i.e., working with $(\delta z)(\boldsymbol{\hat{n}})$ and
$(\delta n)^a(\boldsymbol{\hat{n}})$].  The advantage may
not be so clear in practice, though, given the potentially
limited number of pulsars or stellar sources or their irregular
distribution in the sky.  If the local stochastic background is
dominated by the signal from a handful of nearby sources, then
the background will be non-Gaussian and depart from statistical
isotropy.  This, too, compromises the conceptual advantage of
the $X_{\ell m}$ over the configuration-space description.  For
these reasons, it is beneficial to have at hand also a
description of the correlations in terms of real-space
correlation functions, to which we now turn.

\subsubsection{Correlation functions}
\label{sec:correlations}

The angular two-point autocorrelation function for the redshift
is
\begin{eqnarray}
     C^{zz}(\Theta) &\equiv&\VEV { (\delta
     z)(\boldsymbol{\hat{n}}) (\delta
     z)(\boldsymbol{\hat{m}})}_{\boldsymbol{\hat{n}} \cdot
     \boldsymbol{\hat{m}} =\cos\Theta} \nonumber \\
     &=& \sum_\ell \frac{2\ell+1}{4\pi} C_\ell^{zz} P_\ell(\cos\Theta),
\label{eqn:zzcorrelation}     
\end{eqnarray}
where $P_\ell(\cos\Theta)$ are Legendre polynomials and the angle brackets
denote an average over all pairs of points separated by an angle
$\Theta$.  This (as we will rederive below) is given by the
Hellings-Down curve for an isotropic stochastic background of
transverse-traceless gravitational waves.

Rotationally invariant correlation functions for the angular deflection can be
written in terms of the scalar functions
$E(\boldsymbol{\hat{n}})$ and $B(\boldsymbol{\hat{n}}$),
obtained by taking the divergence and curl,
respectively, of the vector field; these are the correlation
functions $\beta^{EE}(\Theta)$ and $\beta^{BB}(\Theta)$ in
Ref.~\cite{Book:2010pf} and the $EE(\Theta)$ and $BB(\Theta)$
functions in Ref.~\cite{OBeirne:2018slh}.  Although
well-defined mathematically, these scalars can only be computed
from a smooth full-sky map and are unstable to reconstruction
errors.  We therefore work instead (as have prior authors
\cite{Book:2010pf,OBeirne:2018slh,Mihaylov:2018uqm}) with
rotationally invariant correlation functions for vector fields
(following the analogous approach in
Ref.~\cite{Kamionkowski:1996ks} for tensor fields).

Consider the correlation of a vector field $(\delta
n)^a(\boldsymbol{\hat{n}})$ at a point $\boldsymbol{\hat{n}}$ on the sky with a value
$(\delta n)^a(\boldsymbol{\hat{m}})$ at
another point $\boldsymbol{\hat{m}}$.  We can
then consider the great arc connecting these two points on the
sphere and then write these vectors in terms of components
$(\delta n)^{\parallel}$ and $(\delta n)^{\perp}$ that are
parallel and perpendicular, respectively, to that great arc.
There are then two autocorrelations,
\begin{eqnarray}
     C^{\parallel\parallel}(\Theta) &=& \VEV{(\delta
     n)^\parallel(\boldsymbol{\hat{n}}) (\delta
     n)^\parallel(\boldsymbol{\hat{m}})}_{\boldsymbol{\hat{n}}\cdot
     \boldsymbol{\hat{m}} = \cos\Theta}, \nonumber \\
     C^{\perp\perp}(\Theta) &=& \VEV{(\delta n)^\perp(\boldsymbol{\hat{n}})
     (\delta
     n)^\perp(\boldsymbol{\hat{m}})}_{\boldsymbol{\hat{n}}\cdot
     \boldsymbol{\hat{m}}=\cos\Theta},
\end{eqnarray}
and also a cross-correlation,
\begin{equation}
     C^{\perp\parallel}(\Theta) = \VEV{(\delta n)^\perp(\boldsymbol{\hat{n}})
     (\delta n)^\parallel(\boldsymbol{\hat{m}}
     )}_{\boldsymbol{\hat{n}}\cdot  \boldsymbol{\hat{m}} =\cos\Theta},
\end{equation}
that is nonzero only if parity is somehow broken (i.e., if
$C_\ell^{EB} \neq 0$).  There are also two angular
cross-correlation functions,
\begin{eqnarray}
     C^{z\parallel}(\Theta) &=& \VEV{(\delta z)(\boldsymbol{\hat{n}})
     (\delta
     n)^\parallel(\boldsymbol{\hat{m}})}_{\boldsymbol{\hat{n}}\cdot
     \boldsymbol{\hat{m}}=\cos\Theta}, \nonumber \\
     C^{z\perp}(\Theta) &=& \VEV{(\delta z)(\boldsymbol{\hat{n}})
     (\delta
     n)^\perp(\boldsymbol{\hat{m}})}_{\boldsymbol{\hat{n}}\cdot
     \boldsymbol{\hat{m}} =\cos\Theta}, 
\end{eqnarray}
between the redshift and the two components of the deflection
angle aligned with the great arc connecting the points being
correlated.  Again, the latter of these vanishes if there is no
parity breaking (i.e., if $C_\ell^{zB}=0$).  To summarize, there
are, for the combined astrometry/pulsar-timing survey, six
correlation functions ($\parallel\parallel$, $\perp\perp$, $zz$,
$\parallel\perp$, $z\parallel$, and $z\perp$).%

The six sets of correlation functions contain the same
information as the six sets of power spectra.  They are related
to the power spectra through [in addition to
Eq.~(\ref{eqn:zzcorrelation})]
\begin{eqnarray}
     C^{\parallel\parallel}(\Theta) &=& \sum_\ell
     \frac{2\ell+1}{4\pi} \left[C_\ell^{EE} G_{(\ell1)}(\Theta) +
     C_\ell^{BB} G_{(\ell2)}(\Theta) \right], \nonumber\\ \\
     C^{\perp\perp}(\Theta) &=& \sum_\ell
     \frac{2\ell+1}{4\pi} \left[C_\ell^{EE} G_{(\ell2)}(\Theta) +
     C_\ell^{BB} G_{(\ell1)}(\Theta) \right], \nonumber\\
\end{eqnarray}
\begin{eqnarray}
     C^{z\parallel}(\Theta) &=& \sum_\ell
     \sqrt{\frac{2\ell+1}{4\pi}} C_\ell^{zE}
     Y_{(\ell1)}(\Theta,0)  \nonumber \\
     &=& \sum_\ell
     \frac{2\ell+1}{4\pi}\frac{1}{\sqrt{\ell (\ell+1)}} C_\ell^{zE}
     P_\ell^1(\cos\Theta) ,\label{eqn:zparallel} \\
     C^{z\perp}(\Theta) &=& \sum_\ell
     \sqrt{\frac{2\ell+1}{4\pi}} C_\ell^{zB}
     Y_{(\ell1)}(\Theta,0) \nonumber \\
     &=&  \sum_\ell \frac{2\ell+1}{4\pi}
     \frac{1}{\sqrt{\ell (\ell+1)}} C_\ell^{zB}
     P_\ell^1(\cos\Theta), \label{eqn:zperp}
\end{eqnarray}
\begin{equation}
     C^{\perp\parallel}(\Theta) = \sum_\ell \frac{2\ell+1}{4\pi}
     C_\ell^{EB} \left[G_{(\ell1)}(\Theta) - G_{(\ell2)}(\Theta)\right],
\end{equation}     
where
\begin{eqnarray}
     G_{(\ell1)}(\Theta) &\equiv&  -\frac{1}{2} \left[
     \frac{1}{\ell (\ell+1)} P_\ell^2(\cos\Theta) - P_\ell^0(\cos\Theta)
     \right], \nonumber\\
     G_{(\ell2)}(\Theta) &\equiv&  -\frac{1}{\ell (\ell+1)}
     \frac{P_\ell^1(\cos\Theta)}{\sin\Theta},
\end{eqnarray}
and $P_\ell^m(\cos\Theta)$ are associated Legendre polynomials.

The inverse of these relations are
\begin{eqnarray}
     C_\ell^{zz} &=& 2\pi \int_{-1}^1\, d\cos\Theta \, C^{zz}(\Theta)
     P_\ell(\cos\Theta), \\
     C_\ell^{EE} &=& 2\pi \int_{-1}^1 d\cos\Theta\, \left[
     C^{\parallel\parallel}(\Theta) G_{(\ell1)}(\Theta) \right.
     \nonumber \\
     & & \left. -  C^{\perp\perp}(\Theta) G_{(\ell2)}(\Theta) \right],
     \label{eqn:inverserelationEE} \\
     C_\ell^{BB} &=& 2\pi \int_{-1}^1 d\cos\Theta\, \left[
     - C^{\parallel\parallel}(\Theta) G_{(\ell2)}(\Theta)  \right.
     \nonumber \\
     & &    \left.  +     C^{\perp\perp}(\Theta)
     G_{(\ell1)}(\Theta) \right],
     \label{eqn:inverserelationBB}
\end{eqnarray}
\begin{eqnarray}
     C_\ell^{EB} &=& 2 \pi \int_{-1}^1 d\cos\Theta\, C^{\perp\parallel}
     \left[G_{(\ell1)}(\Theta) + G_{(\ell2)}(\Theta)\right] \nonumber\\
     \label{eqn:inverserelationEB}
\end{eqnarray}
\begin{eqnarray}
     C_\ell^{zE} &=& \frac{2\pi}{\sqrt{\ell (\ell+1)}} \int_{-1}^1 d\cos\Theta
     \, C^{z\parallel}(\Theta) P_\ell^1(\cos\Theta),  \nonumber \\
     C_\ell^{zB} &=& \frac{2\pi}{\sqrt{\ell (\ell+1)}} \int_{-1}^1 d\cos\Theta
     \, C^{z\perp}(\Theta) P_\ell^1(\cos\Theta). \nonumber \\
\label{eqn:crossinverse}     
\end{eqnarray} 
Appendix \ref{sec:angularcorrelations} derives these relations.

\section{Predictions for power spectra:  Summary of Results}
\label{sec:summary}

We now provide results for the six power spectra $C_\ell^{XX'}$.
We provide these results for each of the six
possible gravitational-wave polarizations.  As we
will see, all of our results (except for those for the
longitudinal polarization, about which we will say more below)
appear in the form,
\begin{eqnarray}
     C_\ell^{XX',\alpha} &=& 32\pi^2 F_\ell^{X,\alpha}
     \left(F_\ell^{X',\alpha}\right)^* \nonumber \\
     & & \times
     \int\, df\, \frac{6\, H_0^2\, \Omega_\alpha(f)}{(2\pi)^3 f^3}
     W_X(f) W_{X'}(f),
\label{eqn:generalform}     
\end{eqnarray}
where $X$ and $X'$ can be $z$, $E$, or $B$, and the
polarization $\alpha$ can be TE or TB (in general relativity), or
more generally $ST$ or $SL$ (scalar modes), or $VE$ or $VB$
(vector modes).\footnote{Note that statistical homogeneity
requires $\Omega_{TE}(f) = \Omega_{TB}(f)$ and
$\Omega_{VE}(f)=\Omega_{VB}(f)$ \cite{Kamionkowski:2014faa}.
The energy densities in the SL and ST modes are, however, not
required to be the same.}  Here,
the projection factors $F_\ell^{X,\alpha}$ wind
up taking relatively simple forms, summarized in
Table~\ref{tab:F}.  The window functions $W_X(k)$ are related 
to the cadence of observations.  For the simple assumption that
observations are made at two times separated by an interval
$\Delta t$, $W_X(k)=\sin(\pi f \Delta t)$ for all $X$.
More generally, $W_E(k)=W_B(k)$, but $W_z(k)$ (which comes from
different observations) may differ.  We make comments about such
generalizations in Sec.~\ref{sec:timing}.  In the above
equation, $\Omega_\alpha (f)$ is the contribution, per logarithmic
frequency interval, of the type-$\alpha$ gravitational wave to the
critical density, and $H_0$ is the Hubble parameter.

The second line of Eq.~\eqref{eqn:generalform} contributes to the
overall amplitude of the correlation function and incorporates all
frequency dependencies relating to the gravitational wave and the
observation. Omitting this line from Eq.~\eqref{eqn:generalform},
we find the resulting angular correlations functions agree with
previous results. We identify our $C^{\parallel\parallel}(\Theta)$ and
$C^{\perp\perp}(\Theta)$ for the tensor polarizations with
$-\sigma(\Theta)\sin^2\Theta$ and $\alpha(\Theta)\sin^2\Theta$,
respectively, in Ref.~\cite{Book:2010pf}.
We also identify our $C^{\parallel\parallel}(\Theta)$,
$C^{\perp\perp}(\Theta)$, and $C^{z\parallel}(\Theta)$
with $\Gamma_{x\theta}(\Theta)$, $\Gamma_{y\phi}(\Theta)$,
and $\Gamma_{z\theta}(\Theta)$, respectively, for various
polarizations in Ref.~\cite{Mihaylov:2018uqm}.

\begin{table*}
\centering
    \everymath{\displaystyle}
\resizebox{0.8\textwidth}{!}{%
    \begin{tabular}{|c|c|c|c|}
    \hline
	$X$ & $F^{E,X}_\ell$ & $F^{B,X}_\ell$ & $F^{z,X}_\ell$ \\
	\hline \hline
	$ST$ & $\frac{i}{6} \delta_{\ell1}$ & 0 & $-\frac{1}{2\sqrt{2}} \left(\delta_{\ell0} + \frac{i}{3} \delta_{\ell1} \right)$ \\
        	\hline
	$SL$ & $-\frac{i}{3\sqrt{2}} \delta_{\ell1} + \frac{i^{\ell}}{2\sqrt{\ell (\ell+1)}} $ & 0 & $-\frac{i^{\ell}}{4}
        \ln (k r_s)$ \\ \hline
	$VE$ &$\frac{2i}{3\sqrt{2}} \delta_{\ell1} - \frac{i^\ell}{\sqrt{2} \ell (\ell+1)}$ & 0 & $ -\frac{i}{3} \delta_{\ell1} +
        \frac{i^\ell}{\sqrt{2\ell (\ell+1)}}$ \\  \hline
	$VB$ & 0 & $\frac{i}{3\sqrt{2}} \delta_{\ell1} - \frac{i^\ell}{\sqrt{2} \ell (\ell+1)}$ & 0
        \\  \hline
	$TE$ & $-i^{\ell}
        \frac{N_\ell^{-1}}{\sqrt{\ell (\ell+1)}}$ & 0 &
        $\frac{i^{\ell}}{2} N_\ell^{-1}$ \\  \hline
	$TB$ & 0 & $-i^\ell \frac{N_\ell^{-1}}{\sqrt{\ell (\ell+1)}}$
         & 0 \\ \hline     
\end{tabular}
}
\caption{The projection factors that relate the amplitude of a
     given TAM wave to its associated observables.  These also
     determine the power spectra, through
     Eq.~(\protect\ref{eqn:generalform}). The first column $X$
     indicates the gravitational-wave polarization.  Here, $N_\ell
     = \sqrt{(\ell+2)!/[2(\ell-2)!]}$.}
\label{tab:F}
\end{table*}

\section{Calculation of the power spectra and correlation functions}
\label{sec:calcs}

We now calculate the projection factors $F_\ell^{X,\alpha}$, and thus
the power spectra.

\subsection{The redshift}

The redshift $z$ (the fractional frequency shift relative to the
{\it emitted} frequency) of a photon observed from a pulsar at a
distance $r_s$ in a direction $\boldsymbol{\hat{n}}$ in the presence of a
spacetime-metric perturbation $h_{ab}(t,\boldsymbol{x})$ is
(e.g., from Eq.~(23.10) in Ref.~\cite{Maggiorebook}),
\begin{equation}
     z(t,\boldsymbol{\hat{n}}) = \frac12 
     \int_{t-r_s}^t \,dt'\,
     \frac{\partial}{\partial t'} n^a n^b h_{ab}\left(t',\boldsymbol{x}(t')
     \right),
\end{equation}
where $\partial/\partial t'$ acts only on the first argument,
and not the time dependence in $\boldsymbol{x}(t')$. 

Now consider a perturbation,
\begin{equation}
     h_{ab}(\boldsymbol{x},t) = 4\pi i^\ell h_{k\ell m}^\alpha
     \Psi^{k,\alpha}_{(\ell m)ab}(\boldsymbol{x}) e^{-ikt},
\end{equation}
due to a single TAM wave of polarization $\alpha$ with amplitude
$h_{k \ell m}^\alpha$.  We assume, with the $e^{-ikt}$ time
dependence, that the waves propagate at the speed of light with
angular frequency $\omega=k$.  For the
calculation of the redshift, we need the quantity $n^a n^b
\Psi^{k,\alpha}_{(\ell m)ab}(\boldsymbol{x})$, which can be written as a
spherical harmonic $Y_{\ell m}(\boldsymbol{\hat{n}})$ times some radial function
$-R_\ell^{L,\alpha}(kr)$ provided in Appendix \ref{sec:TAM}.  For
example, the radial function
for TE is $R_\ell^{L,TE}(kr) = -N_\ell j_\ell(kr)/(kr)^2$, where
$N_\ell = \sqrt{(\ell+2)!/[2(\ell-2)!]}$ and $j_\ell(kr)$ is the
spherical Bessel function, and it
vanishes for VB and TB.   We then find that the redshift $z$ due
to a single $\alpha$ mode is
\begin{equation}
     z(\boldsymbol{\hat{n}},t) = -\frac{i}{2}
     4\pi i^\ell h_{k\ell m}^{\alpha}
     Y_{\ell m}(\boldsymbol{\hat{n}}) e^{-ikt}
     \int_0^{k r_s}\, dx\, R_\ell^{L,\alpha}(x) e^{ix},
\label{eqn:redshift}
\end{equation}
with $x=kr$. For example, for TE, this is
\begin{equation}
     z(\boldsymbol{\hat{n}},t) = \frac{i}{2}
     4\pi i^\ell h_{k\ell m}^{TE}
     N_\ell Y_{\ell m}(\boldsymbol{\hat{n}}) e^{-ikt}
     \int_0^{k r_s}\, dx\, \frac{j_\ell(x) }{x^2} e^{ix}.
\end{equation}
We then take the distant-source limit $k r_s \to \infty$ and
thus infer that
\begin{equation}
     z(\boldsymbol{\hat{n}},t) = 4\pi i^\ell F_\ell^{z,\alpha} h_{k\ell m}^\alpha Y_{\ell m}(\boldsymbol{\hat{n}}) e^{-ikt},
\label{eqn:redshiftresult}
\end{equation}
with
\begin{equation}
     F_\ell^{z,\alpha} = -\frac{i}{2} \int_0^\infty \, dx\,
     R_\ell^{L,\alpha}(x) e^{ix}.
\end{equation}
For example, for the TE mode, the integral evaluates to
\begin{equation}
     F_\ell^{z,TE} = \frac{i^{\ell}}{2} N_\ell^{-1}.
\end{equation}
The analogous results for the other five polarizations are
provided in Table~\ref{tab:F}.

The pulsar-timing spherical-harmonic coefficient for the
observable change in the pulsar frequency, due to mode $\alpha$,
is then  obtained, differencing the result at two different
times separated by $\Delta t$, by
\begin{equation}
     z_{\ell m}(t) = 4\pi i^\ell F_\ell^{z,\alpha} h_{k\ell m}^E \left(e^{-i k
     \Delta t}-1 \right) e^{-ikt},
\end{equation}
where here $t$ is the time of the initial observation.
This particular TAM wave then contributes
\begin{equation}
     \left(C_\ell^{zz}\right)_{k\ell m} = (4\pi)^2 \left| F_\ell^{z,\alpha}
     \right|^2 \left| h_{k\ell m}^E
     \right|^2 \,\left[ 2W(k) \right]^2,
\end{equation}
to the power spectrum for the redshift observable.  Here, $W(k)
\equiv \sin (k \Delta t/2)$ is the frequency-space window
function.

Now suppose we have a stochastic background characterized by a
power spectrum $P_h(k)$, using the conventions/definitions of
Sec. V.A in Ref.~\cite{Dai:2012bc}.  We then infer that each
spherical-harmonic coefficient $z_{\ell m}$ takes on a value
selected from a distribution with zero mean and variance
$\VEV{\left|z_{\ell m}\right|^2} = C_\ell^{zz}$, where
the angular power spectrum $C_\ell^{zz}$ is obtained by summing
over all $\alpha$-mode TAM waves with the same TAM quantum
numbers $\ell m$.  Thus,
\begin{equation}
     C_\ell^{zz,\alpha} = \sum_k
     \left(C_\ell^{zz}\right)_{k\ell m}
     = \frac{4}{\pi} \left|F_\ell^{z,\alpha} \right|^2
     \int\, k^2 \, dk\, P_h(k) \left[ W(k) \right]^2.
\label{eqn:zTEresult}     
\end{equation}
We then use Eq.~(\ref{eqn:translation}) to recover the form
given in Eq.~(\ref{eqn:generalform}).

\subsubsection{Specific results}

\begin{figure}[htbp]
\centering
\includegraphics[width=0.9\linewidth]{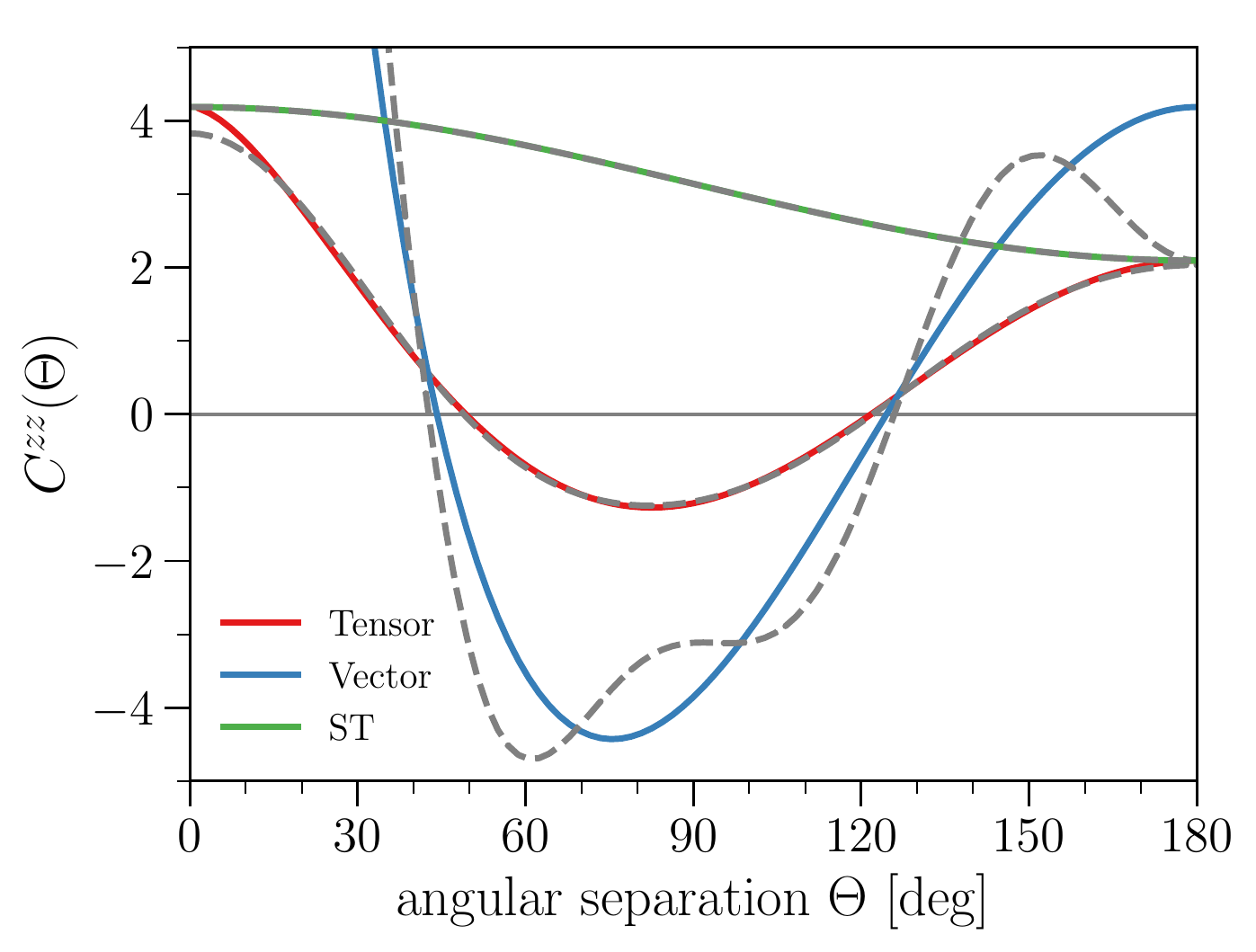}
\caption{The $C^{zz}(\Theta)$ correlation
     functions for the transverse-traceless tensor modes,
     vector modes, and the ST mode.
     They are normalized
     by omitting second line in Eq.~\eqref{eqn:generalform}.
     The solid curves show the exact results in the
     distant-source limit, and dashed curves show the
     results from truncating the multipole expansion
     at $\ell_\textrm{max}=5$.}
\label{fig:zzcorrelations}
\end{figure}

{\it Transverse-traceless modes.}
We begin  with the transverse-traceless modes that propagate
in general relativity.  The results in this case are obtained
exclusively from the TE modes, since TB does not contribute to
the redshift.  We obtain from Eq.~(\ref{eqn:zTEresult}),
\begin{equation}
     C_\ell^{zz,GW} = \frac{12 \,H_0^2 N_\ell^{-2}}{\pi} \int\, df\,
     \frac{\Omega_{GW}(f)}{f^3} \left|W_z(f)\right|^2,
\label{eqn:zGWresult}     
\end{equation}
where we used $\Omega_{TE}(f)=\Omega_{GW}(f)/2$, and
$\Omega_{GW}(f)$ is the gravitational-wave energy density
(summing over both polarization states).  Since
$C_\ell \propto \ell^{-4}$ at larger $\ell$, the power spectrum is very
highly peaked at the smallest multipole moments, and
particularly the quadrupole.  The $\ell$ dependence of the power
spectrum is the same for any functional form of
$\Omega_{GW}(f)$, a consequence of the distant-source
limit---the observations probe the {\it local} spacetime-metric
perturbation.  Using the results of
Appendix~\ref{sec:legendretricks}, the angular correlation
function is found, for the canonical transverse-traceless modes,
to be
\begin{equation}
     C^{zz,GW}(\Theta) = \frac{3 \,H_0^2 }{2\pi^2} \int\, df\,
     \frac{\Omega_{GW}(f)}{f^3} \left|W_z(f)\right|^2\,
     \textrm{HD}(\theta),
\end{equation}
where $\textrm{HD}(\Theta)$ is the famous Hellings-Downs curve provided
in Eq.~(\ref{eqn:HD}).
The angular correlation function is shown in Fig.~\ref{fig:zzcorrelations}.

{\it Vector modes.}  The redshift power spectrum for the vector
modes is exactly as in Eq.~(\ref{eqn:zGWresult}), but with
$N_\ell^{-2} \to [2\ell (\ell+1)]^{-1} -(2/9)\delta_{\ell1}$.  Simple
algebraic manipulation of results in
Appendix~\ref{sec:legendretricks} yields the vector analog,
\begin{eqnarray}
     \textrm{HD}_v(\Theta) &=& 2 \sum_{\ell=1}^\infty
     (2\ell+1)\left(\frac{1}{2\ell (\ell+1)} -\frac{2}{9}\delta_{\ell1}
     \right) P_\ell(\cos\Theta) \nonumber \\
     &=& -2\ln \left[\sin(\Theta/2)
     \right]-1 -\frac{4}{3}\cos\Theta,
\end{eqnarray}
which is shown in Fig.~\ref{fig:zzcorrelations} and agrees with results obtained from real-space calculations
\cite{Lee:2008aa,Mihaylov:2018uqm}.  The logarithmic
divergence as $\Theta\to 0$ arises in the harmonic approach
given that the summand is $\sim \ell^{-1}$ at large $\ell$.
This divergence is regulated by taking $kr_s$ finite.  In
practice, the divergence is irrelevant given the
finite density of pulsars on the sky.

{\it Scalar-transverse modes.}  The power spectrum is again as
in Eq.~(\ref{eqn:zGWresult}), but now with 
$N_\ell^{-2} \to (\delta_{\ell0}+\delta_{\ell1}/9)/8$.  The
Hellings-Downs analogue then becomes simply $1/4+(1/12)\cos\Theta$,
as shown in Fig.~\ref{fig:zzcorrelations} and again in agreement with prior work \cite{Lee:2008aa}.  In
principle, the monopole would be observable if we had a
complete handle on timing information from a terrestrial standard clock.
In practice, though, errors in timing and timing models can
produce monopolar correlations between pulsars~\cite{2012MNRAS.427.2780H},
rendering the extraction of the monopole difficult.
There is also no cross-correlation with angular deflections,
since there is no monopole for angular deflections.

{\it Scalar-longitudinal.}  The radial function $R_\ell^{L,SL}$
for the SL mode contains a term $\propto j_\ell(x)$ that renders
the radial integral divergent in the distant-source limit $kr_s
\to \infty$.  This is a consequence of the fact that the light
ray from a source aligned with the direction of propagation of a
gravitational wave can ``surf'' the gravitational wave and (unlike
the other modes) experiences a stretching in this same
direction.  The magnitude of the redshift thus accrues
monotonically as the light ray propagates from the source.
The integral can be performed numerically (or written in terms of
hypergeometric functions, which are then determined
numerically), but can, using $j_\ell(x) \sim x^{-1}
\cos(x-(\ell+1)\pi/2)$ for $x\gg \ell$, be approximated
in the $kr_s \gg 1$ limit by
\begin{equation}
     F_\ell^{z,SL}(k) = - \frac{i^{\ell}}{4} \ln(kr_s).
\label{eqn:logarithm}     
\end{equation}
Note that this result, unlike all the others we encounter in
this paper, depends on the wave number $k$ and on the source
distance $r_s$.  It is also, strictly speaking, valid only for
$kr_s \gg \ell$.  Given the logarithmic dependence on both $k$ and $r_s$, we
can obtain rough estimates by fixing the logarithm using some
characteristic $k$ [set, perhaps by the observationally
preferred frequency $f=k/(2\pi)\simeq$yr$^{-1}$] and source
distance (perhaps $\sim$3 kpc).  With these canonical values $k
r_s\sim6\times 10^4$ (justifying the $kr_s \gg \ell$ assumption),
and the logarithm is roughly 10, explaining the roughly
order-of-magnitude enhancement inferred numerically in previous
work \cite{Mihaylov:2018uqm,OBeirne:2018slh}.  
Since the logarithm grows very slowly, the asymptotic
expression in Eq.~(\ref{eqn:logarithm}) is unlikely to be
numerically precise, possibly with significant contributions
from subdominant terms.

The multipole-moment ($\ell$) dependence of the power spectrum is
also interesting.  In the distant-source limit, it is
independent of $\ell$.  Such a power spectrum is that for white
noise, which exhibits a correlation function that is nonzero only
at zero lag (formally, a Dirac delta function).  This may
account for numerical evidence for a rapid increase of
$C^{zz}(\Theta)$ as $\Theta\to0$ for the SL mode.
Phenomenologically, it implies that the SL mode gives rise to
fluctuations that are uncorrelated from one point on the sky to
the other.  Since the large-$x$ approximation for
$j_\ell(x)$ used to obtain Eq.~(\ref{eqn:logarithm}) breaks down
for $\ell \gtrsim kr_s\sim 6\times 10^4$, we surmise that the
correlation should be nonzero at angular separations $\Theta
\lesssim 180^\circ/\ell \simeq 10$~arcsec.

We quantify these statements by augmenting the SL projection factor with a Gaussian in $\ell$, to $F_\ell = -(i^{\ell}/4) \ln( k r_s) e^{-\ell^2/2 \ell_{\textrm{max}}^2}$, to account for the breakdown in the distant-source limit at $\ell \gtrsim \ell_{\rm max}$.  With this, the ``Hellings-Downs" curve for the SL modes becomes
\begin{align}
    \textrm{HD}_{SL}(\Theta)
    &= \frac14 \sum_{\ell=0} (2\ell+1) P_\ell(\cos\theta)
    e^{-\ell^2/\ell_{\textrm{max}}^2} \nonumber\\
    & \simeq \frac{\ell_{\textrm{max}}^2}{4} e^{-\ell_{\rm max}^2 \theta^2}.
\end{align}

\subsection{Angular deflections}

As derived in prior work~\cite{Book:2010pf}, the angular deflection of a light ray
observed at time $t$ propagating in the $\boldsymbol{\hat{n}}$ direction from
a source at distance $r_s$ is
\begin{widetext}
\begin{equation}
     (\delta n)^a (\boldsymbol{\hat{n}},t) = \Pi^{ac} n^b \left\{-\frac12 
     h_{bc}(t,\boldsymbol{0})+ \frac{1}{r_s} \int_0^{r_s} \, dr\, \left[
     h_{bc}(t-r,r\boldsymbol{\hat{n}})  - \frac{r_s-r}{2} n^d\partial_c
     h_{bd}(t-r,r\boldsymbol{\hat{n}}) \right] \right\},
\end{equation}
where $\Pi_{ab}(\boldsymbol{\hat{n}})=g_{ab}-\hat{n}_a \hat{n}_b$ projects onto
the plane orthogonal to $\boldsymbol{\hat{n}}$ (i.e., onto the plane of the
sky).  Since we are not concerned with sources at cosmological distances,
we take the spacetime metric $g_{ab}$ to be Minkowski.
Using the relation \cite{Dai:2012bc},
\begin{equation}
     \hat{n}^b \hat n^d \partial_c h_{bd} = \partial_c(\hat n^b \hat
     n^d h_{bd}) - \frac{2}{kr} \Pi_{cb} \hat n_d h^{bd},
\label{eqn:nabla}
\end{equation}
the angular deflection can be rewritten,
\begin{equation}
     (\delta n)^a (\boldsymbol{\hat{n}},t)= \Pi^{ac} \left\{-\frac12 
     n^b h_{bc}(t,\boldsymbol{0})+ \int_0^{r_s} \, dr\, \left[
     \frac{1}{r} n^b h_{bc}(t-r,r\boldsymbol{\hat{n}}) - \frac{r_s-r}{2 r_s}
     \partial_c n^b n^d  h_{bd}(t-r,r\boldsymbol{\hat{n}}) \right] \right\}.
\label{eqn:deflection}
\end{equation}
\end{widetext}

Now consider a single TAM wave of polarization $\alpha$, quantum
numbers $k\ell m$, and amplitude $h^{\alpha}_{k\ell m}$.  The first
term in Eq.~(\ref{eqn:deflection}), the ``observer'' term, is
obtained by evaluating the coefficients of
$Y_{(\ell m)a}^E(\boldsymbol{\hat{n}})$ and
$Y_{(\ell m)a}^B(\boldsymbol{\hat{n}})$ in
Eq.~(\ref{eqn:centralresult}) at $r=0$ (the projection operator
$\Pi^{ac}$ does not affect the E and B vector spherical harmonics,
since they are already defined on the 2-sphere of the sky).
These turn out to be
nonzero only for $\ell=2$ and only for the ST, SL, VE, and TE
coefficients of $Y_{(\ell m)a}^E(\boldsymbol{\hat{n}})$.  As a
result the first term in Eq.~(\ref{eqn:deflection}) evaluates to
$F_\ell^{E,\alpha (0)} Y_{(\ell m)a}^E(\boldsymbol{\hat{n}})$,
with $F_\ell^{E,\alpha (0)}=c\delta_{\ell 2}$ and
$c=-\sqrt{6}/30$ for SL, $c=\sqrt{3}/30$ for ST, and
$c=(5\sqrt{2})^{-1}$ for VE and TE.

The second term in Eq.~(\ref{eqn:deflection}) (the first term in
the integral) receives contributions from all terms in
Eq.~(\ref{eqn:centralresult}).
The contribution from these terms
to the angular deflection is
\begin{equation}
     F_\ell^{X,\alpha(1)} = \int_0^{k r_s} \, \frac{dx}{x}
     e^{ix} R_\ell^{X,\alpha}(x),
\end{equation}
where the radial function $R_\ell^{X,\alpha}(x)$ is the coefficient of
the appropriate $Y_{(\ell m)a}^X(\boldsymbol{\hat{n}})$ in
Eq.~(\ref{eqn:centralresult}), and $X=\{E,B\}$.
The integrals are all finite and
easily evaluated in the distant-source limit $kr_s \to \infty$.

The last term in Eq.~(\ref{eqn:deflection}) receives, as
discussed at the end of Appendix \ref{sec:TAM}, contributions
only from the $Y_{(\ell m)a}^L(\boldsymbol{\hat{n}})$ terms in
Eq.~(\ref{eqn:centralresult}).  The evaluation of this term is
then aided by the relation  $\Pi_{ab} \nabla^b Y_{(\ell
m)}(\boldsymbol{\hat{n}}) = -
M_{\perp a} Y_{(\ell m)}(\boldsymbol{\hat{n}}) / r = -
\sqrt{\ell (\ell+1)} Y^E_{(\ell m)a}(\boldsymbol{\hat{n}}) / r$, 
where $M_{\perp a}$ is the gradient operator on the sphere
\cite{Dai:2012bc}.  The contributions from these terms to the
angular deflection are
\begin{equation}
     F_\ell^{E,\alpha(2)} = - \frac{\sqrt{l(l+1)}}{2}
     \int_0^{k r_s} \, \frac{dx}{x} e^{ix}
     \frac{k r_s -x}{kr_s} R_\ell^{L,\alpha}(x).
\end{equation}
The integrals are again all finite and easily evaluated in the
distant-source limit $kr_s \to \infty$.

Putting the results together, the angular deflection from this
TAM mode is
\begin{align}
    (\delta n)_a(\boldsymbol{\hat{n}}) =
    4&\pi i^\ell h^\alpha_{k\ell m} e^{-ikt} \times \nonumber\\
    &\left[F_\ell^{E,\alpha} Y^E_{(\ell m)a}(\boldsymbol{\hat{n}})
      +F_\ell^{B,\alpha} Y^B_{(\ell m)a}(\boldsymbol{\hat{n}}) \right],
\end{align}
where the $F_\ell^{E,\alpha}$ and $F_\ell^{B,\alpha}$ are the sums of
the three individual contributions and listed in
Table~\ref{tab:F}.  Interestingly, the observer terms for
$F_\ell^{E,\alpha}$ augment the radial-integral contributions that
arise for $\ell=2$, yielding very compact expressions
in the table. The corresponding power spectra, as given in
Eq.~(\ref{eqn:generalform}), are then obtained, following the
same steps as above for the redshift, by taking the difference
between the angular deflections evaluated at two different times
separated by $\Delta t$, and then squaring and then summing over
all wave numbers $k$ for a given $\ell m$.

\begin{figure}[htbp]
\centering
\includegraphics[width=0.9\linewidth]{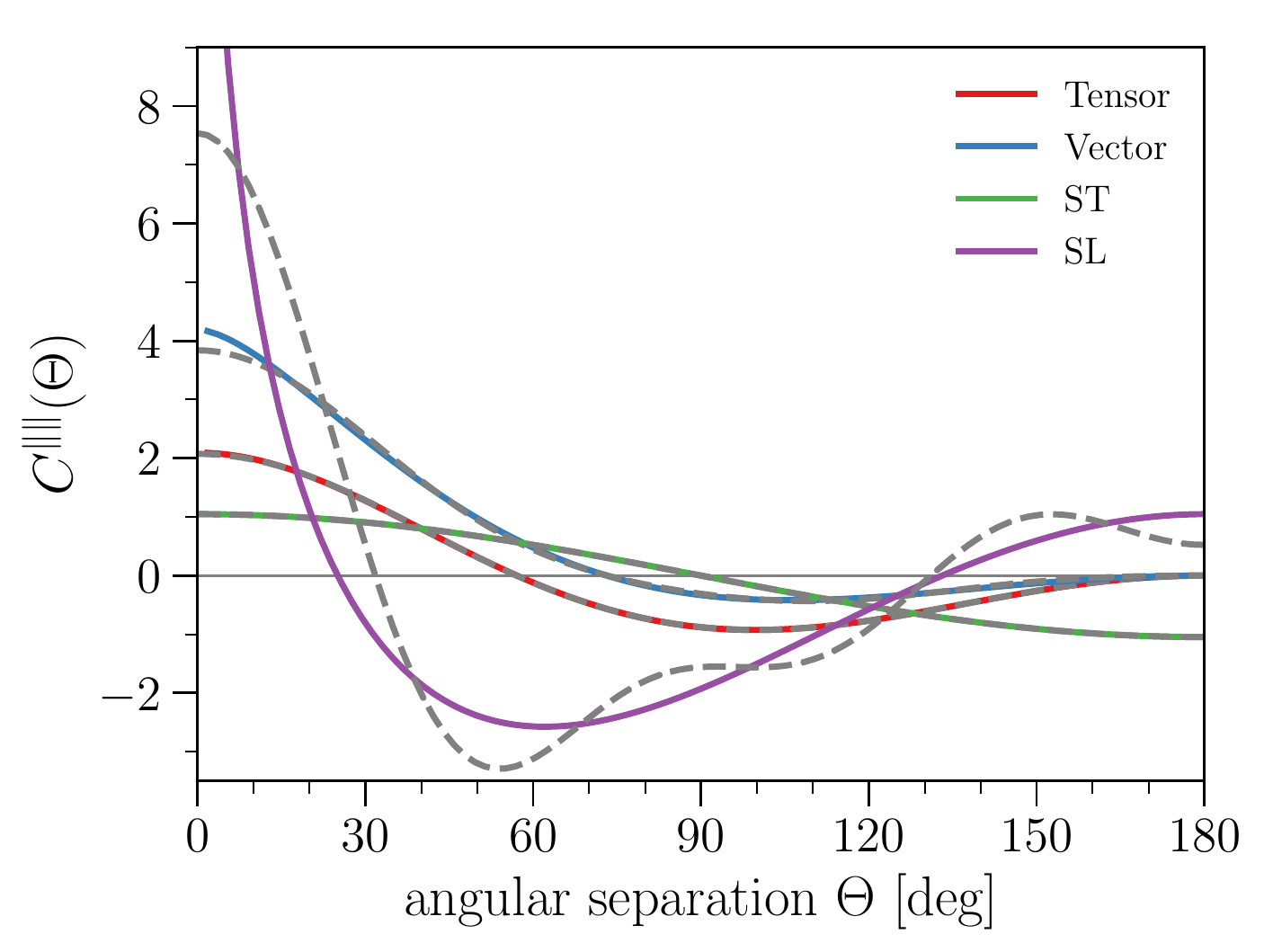}
\caption{The $C^{\parallel\parallel}(\Theta)$ correlation
     functions for the transverse-traceless tensor modes,
     vector modes, and the ST and
     SL modes.  They are normalized
     by omitting second line in Eq.~\eqref{eqn:generalform}.
     The solid curves show the exact results in the
     distant-source limit, and dashed curves show the
     results from truncating the multipole expansion
     at $\ell_\textrm{max}=5$.}
\label{fig:angularcorrelations}
\end{figure}

\begin{figure}[htbp]
\centering
\includegraphics[width=0.9\linewidth]{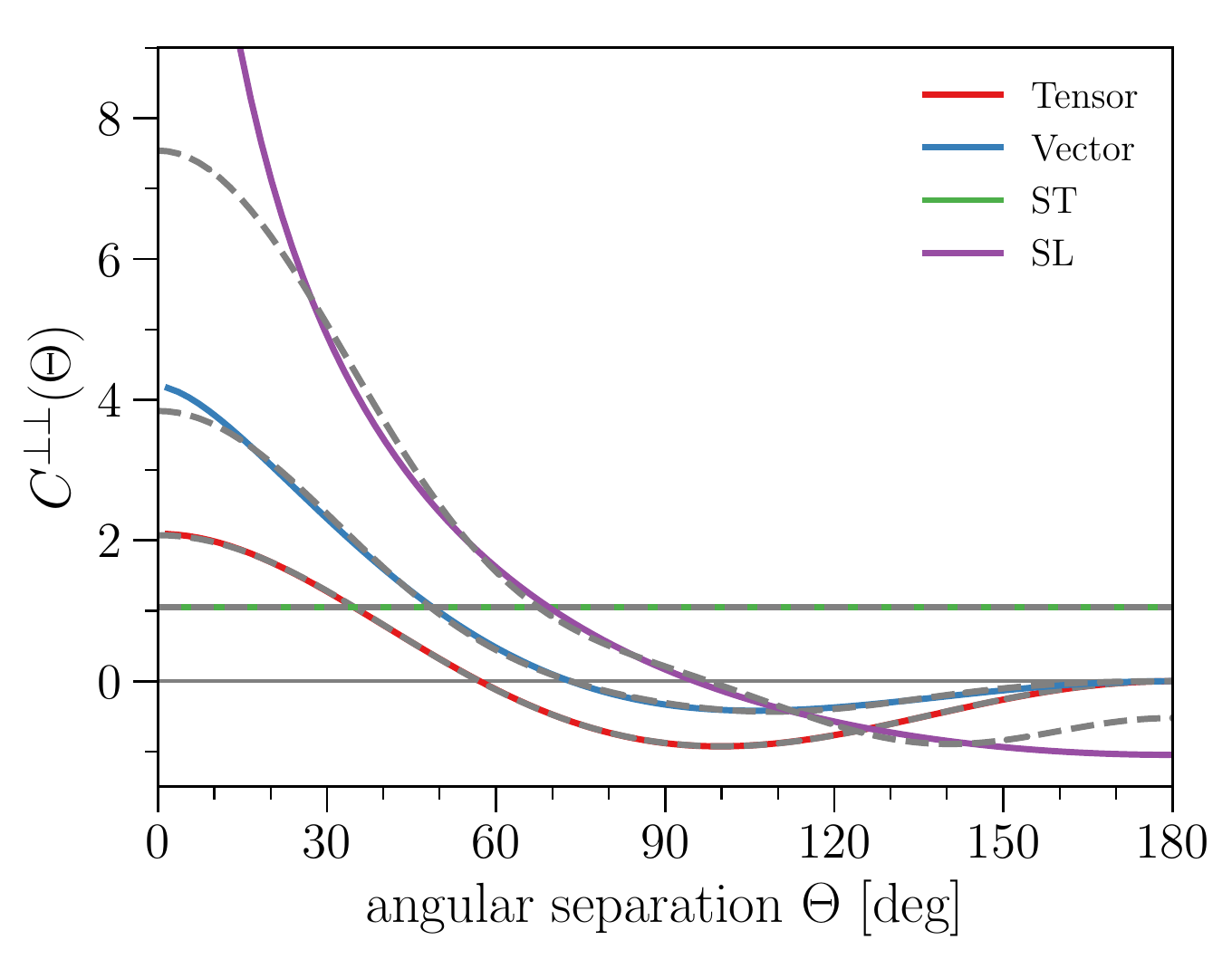}
\caption{The same as Fig.~\ref{fig:angularcorrelations},
     but for the $C^{\perp\perp}(\Theta)$ correlation
     functions. The correlation functions are the same
     as $C^{\parallel\parallel}(\Theta)$ for the vector
     and tensor modes.}
\label{fig:perp}     
\end{figure}

\begin{figure}[htbp]
\centering
\includegraphics[width=0.9\linewidth]{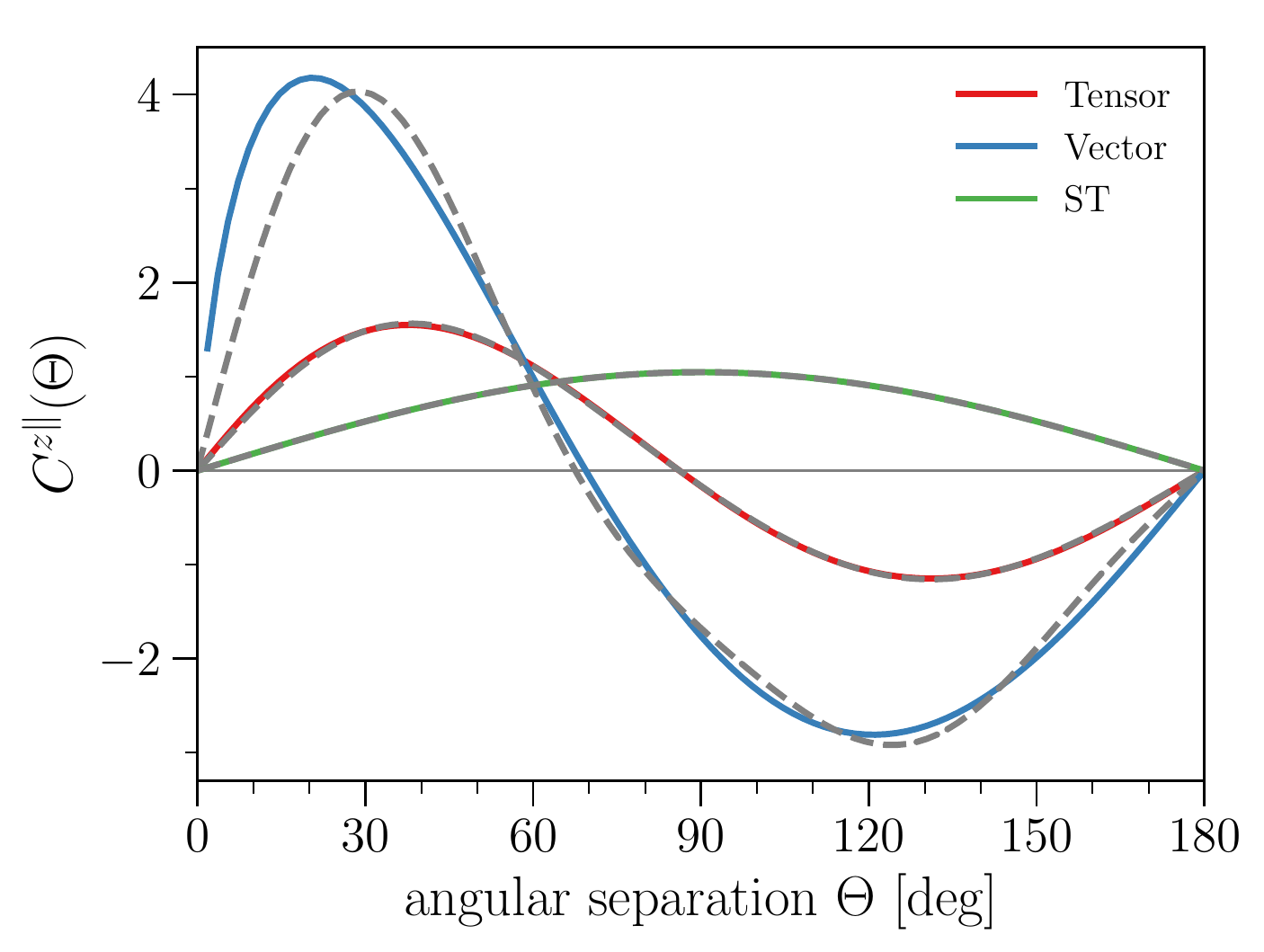}
\caption{The $C^{z\parallel}(\Theta)$ correlation
     functions for the transverse-traceless tensor modes,
     vector modes, and the scalar-transverse (ST).
     They are normalized
     by omitting second line in Eq.~\eqref{eqn:generalform}.
     The solid curves show the exact results in the
     distant-source limit, and dashed curves show the
     results from truncating the multipole expansion
     at $\ell_\textrm{max}=5$.}
\label{fig:crosscorrelations}
\end{figure}

\subsubsection{Specific results}

{\it Transverse-traceless modes.}  The power spectra for the
gravitational waves that appear in general relativity are
\begin{eqnarray}
     C_\ell^{EE,GW} &=& C_\ell^{BB,GW} \nonumber\\
     &=& \frac{12 \,H_0^2 N_\ell^{-2}}{\pi
     \ell (\ell+1)} \int\, df\,
     \frac{\Omega_{GW}(f)}{f^3} \left|W_z(f)\right|^2,
\end{eqnarray}
and we note that this is equal to $[\ell (\ell+1)]^{-1}
C_\ell^{zz,GW}$.  As a
result, the correlation functions 
$\beta^{EE}(\Theta)=\beta^{BB}(\Theta) \propto
\textrm{HD}(\Theta)$, as noted previously.\footnote{An
explanation of this coincidence is provided in Appendix
\protect\ref{sec:HDsimple}.}  The
$\parallel\parallel$ and $\perp\perp$ correlation functions are
easily evaluated numerically and shown in
Fig.~\ref{fig:angularcorrelations}.  Although they are nominally
obtained from an infinite sum, numerically precise results can
be obtained from just the first few terms, given the steep drop
of $C_\ell$ with $\ell$, as seen in
Figs.~\ref{fig:angularcorrelations} and \ref{fig:perp}.
We have checked numerically that these
correlation functions agree with those in prior work.  They can
also be shown analytically to agree by writing the associated
Legendre polynomials $P_\ell^2(x)$ and $P_\ell^1(x)$ in the
definitions of $G_{(\ell1)}(x)$ and $G_{(\ell2)}(x)$ in terms of
Legendre polynomials and then using the techniques of Appendix
\ref{sec:legendretricks}.  The derivation is straightforward
but not particularly illuminating, and so we leave out the details.

{\it Vector modes.}  The EE/BB power spectra for these
modes are again equal and turn out to be
$C_\ell^{EE,\textrm{vector}} = C_\ell^{BB,\textrm{vector}}
\propto [\ell (\ell+1)]^{-2}$ for $\ell >1$, with an additional
contribution (that is the same for EE and BB) for $\ell=1$.  The
E/B correlation functions
$\beta^{EE}(\Theta) = \beta^{BB}(\Theta)=\textrm{HD}_v(\Theta)$
in this case turn out to be the same as the angular redshift
correlation (which can again be understood simply from the
arguments in Appendix \ref{sec:HDsimple}).  Again, the
rotationally invariant angular correlation functions are shown in
Figs.~\ref{fig:angularcorrelations} and \ref{fig:perp} and agree with those in
Refs.~\cite{OBeirne:2018slh,Mihaylov:2018uqm}.

{\it Scalar modes.}  Statistical homogeneity implies equal TE
and TB powers for transverse-traceless modes and equal VE and VB
powers for vector modes.  There is, however, no corresponding
symmetry requirement that the SL and ST modes should have the
same power \cite{Yunes:2013dva,Isi:2018miq}.  The relative amplitudes may
depend on the details of the alternative-gravity theory.%
\footnote{We reiterate that the calculations in this work
assume gravitational waves propagate at the speed of light; if
this is not the case, the modified dispersion relation
$\omega(k)$ must be used in the expression for plane-wave
propagation $e^{-i\omega(k)t}$.}  For
example, Brans-Dicke theory has a massless scalar that excites
the ST mode \cite{Eardley:1974nw}.  In $f(R) = R + \alpha R^2$
gravity, there is a single massive scalar mode that introduces a
mixture of SL and ST modes, with a ratio dependent on $\alpha$
\cite{Liang:2017ahj}.  In the more general case of Horndeski
gravity, the trend is the same: a massless scalar mode excites
ST modes, while a massive scalar excites both SL and ST
modes \cite{Hou:2017bqj}.  The two modes must therefore be
considered separately.

The EE correlation function $\beta^{EE}(\Theta)$ for the SL mode
exhibits a Dirac delta function with an added dipole, and for
the ST mode it is a pure dipole.  The B-mode correlation
$\beta^{BB}(\Theta)=0$ for both scalar modes.  The
rotationally invariant angular correlation functions are shown in
Figs.~\ref{fig:angularcorrelations} and \ref{fig:perp}; note
that $C^{\parallel\parallel}(\Theta)$ and
$C^{\perp\perp}(\Theta)$ are unequal for these modes.  The
angular correlations for the ST modes are simply
$C^{\parallel\parallel} =(3/8\pi) C_{\ell=1}^{EE}\cos\Theta$
and $C^{\perp\perp} =(3/8\pi) C_{\ell=1}^{EE}$.

\section{Redshift-deflection cross-correlation}
\label{sec:crosscorrelation}

Since the redshift and E-mode deflection angle both
arise from the same TE TAM waves, there is a cross-correlation
between these two observables characterized by a power spectrum,
\begin{equation}
     C_\ell^{zE} = \VEV{z_{\ell m} E_{\ell m}^*}.
\end{equation}
Moreover, since the $k$ integrands in the expressions for the
E-mode and redshift power spectra are identical, this
cross-correlation is {\it exact} (for concurrent PTA/astrometry
observations) in the distant-source limit; i.e.,
\begin{equation}
     C_\ell^{zE} = \sqrt{C_\ell^{EE} C_\ell^{zz}},
\end{equation}
(except for the SL mode, for which the redshift
is sensitive to the location of the source, while the
deflection is dominated by the local metric perturbations,
resulting in essentially no cross-correlation).
As a result, pulsar-timing and astrometry probes of the
stochastic gravitational-wave background can be used to
cross-check.  This exact cross-correlation moreover suggests
that {\it astrometry and pulsar-timing surveys can be used to
complement each other to optimize sky coverage in the event that
there are blind regions in the sky in one survey or the other.}

These cross-correlations can be obtained numerically and are
provided in Fig.~\ref{fig:crosscorrelations}.  Again, the
correlations are well characterized by the lowest multipole
moments.  The analytic results for $C^{z\parallel}(\Theta)$ in
Eqs.~(53) and (55) of Ref.~\cite{Mihaylov:2018uqm}, for tensor
and vector modes, respectively, can be obtained from the 
relation,
\begin{equation}
     P_\ell^1(\cos\Theta) = \frac{\ell(\ell+1)}{2\ell+1}
     \frac{P_{\ell+1}(\cos\Theta) - P_{\ell-1}(\cos\Theta)}{\sin\Theta},
\end{equation}
and those in Appendix~\ref{sec:legendretricks}.

\section{Gravitational-wave window functions, pointing, and chirality}
\label{sec:windowfns}

\subsection{Time evolution and window functions}
\label{sec:timing}

We now note that all of the predicted power spectra can be
written as the product of a function of multipole moment $\ell$
and an integral,
\begin{equation}
     \int\, k^2 \, dk\,P_h(k) \left[W(k;\Delta t) \right]^2,
\label{eqn:structure}     
\end{equation}
for the power spectrum (with analogous results for vector and
scalar modes).  The absence of any dependence of the angular
structure on the form of $P_h(k)$ arises because the observables
arise only in the distant-source limit.  We probe with
these measurements only the {\it local} spacetime-metric
perturbation; there are no long-range spatial correlations
imprinted on the observed angular correlations.

We now focus on the window function $W(k;\Delta t) =\sin(k\Delta
t/2)$ obtained by assuming redshifts and stellar positions were
obtained at two instantaneous times separated by an interval
$\Delta t$.  The window function is then just the Fourier
transform of a time ``exposure,'' $\delta_D(t+\Delta t) -
\delta_D(t)$.  More realistically, the measurements may be done
over some range of times, or (for pulsar timing) inferred from
timing residuals.  The detailed functional form of the window function
$W(k)$ will therefore differ from the simple $W(k;\Delta t)$
inferred here.  Regardless, we expect $[W(k)]^2 \propto (k
\Delta t)^2$ for $k\Delta t\ll 1$, where $\Delta t \sim$yrs is
the overall time interval in which measurements are done.  Also,
there will be a suppression at high $k$ that arises from the
finite duration of any particular measurement made.

As discussed in Sec. \ref{sec:crosscorrelation} above, the
angular cross-correlation between the astrometry and
pulsar-timing signals are {\it exact} if the two observing
periods coincide.  More generally, though, the observations will
not necessarily be concurrent, and so the astrometry--PTA
cross-correlation will be degraded.  For example, suppose the
pulsar-timing measurements are done at two times $t$ and
$t+\Delta t$, while the astrometry measurements are done at
times $t+\delta t$ and $t+\delta t +\Delta t$.  The
cross-correlation coefficient will then be
\begin{eqnarray}
     r &\equiv& \frac{ \VEV{ z_{lm} E_{lm}^*}}{
     \sqrt{\ \VEV{\left|z_{lm} \right|^2} \VEV{ \left|E_{lm}
     \right|^2}}} \nonumber \\
     &=& \frac{ \int \frac{dk}{2\pi} P(k) \sin^2(k\Delta t/2)
     \cos (k\delta t/2)}{ \int \frac{dk}{2\pi} P(k)
     \sin^2(k\Delta t/2)}.
\end{eqnarray}  
One can see that if the PTA and astrometry measurements are
separated by times $\delta t \lesssim \Delta t$, the
cross-correlation remains strong and then becomes weak for
$\delta t \gtrsim \Delta t$.

Here we have assumed that pulsar-timing and astrometry results
are each made at only two epochs.  In this case, the
measured spherical-harmonic coefficients for each $\ell m$ receive
contributions from an array of TAM waves $h_{k\ell m}$ for an array
of values of wave number $k$, and if the observations are
concurrent, the {\it same} set of TAM waves.  If, however,
measurements are made over a larger set of times---say $N$
times, rather than two---then the measurements can be decomposed
into power spectra for $N-1$ different window functions, which
probe different ranges of frequencies.  If so, then information
about the distribution of the wave numbers $k$ that give rise to
the signal, for each $\ell m$, can be extracted.  In other words,
with time measurements, {\it the three-dimensional spacetime-metric
perturbation (and not just some two-dimensional projection) can
begin to be reconstructed}.  We leave an elaboration of this
frequency-space analysis for future work.

Suppose now that the stochastic background has an energy
density $\Omega_{\rm gw}(f) \sim$~constant, which is expected
for the nearly scale-invariant spectrum ($n_t\simeq0$) generated
from inflation.  In this case, the window-function behavior
$W(k)\propto k$ for $k \Delta t \ll 1$ results in an equal 
contribution per logarithmic frequency interval (at frequencies
$k \Delta t \ll 1$) to the observables.  If so, then the
distant-source limit we have employed is not strictly speaking
valid.  We have checked (but leave details for elsewhere), that
the contribution of longer-wavelength modes (i.e., those with $k
r_s \lesssim 1$) to $\ell\geq3$ multipole moments is suppressed
relative to what is inferred using the distant-source limit.
The contribution to the quadrupole is, however, a bit larger.
Still, given that the amplitude of the inflationary background
is expected to be far too small ($\Omega_{\rm gw} h^2 \sim 10^{-16}$
at frequencies $f\sim 10^{-9}$~Hz) to be accessed with PTAs and
astrometry~\cite{Caprini:2018mtu}, we consider this point academic.

The background more realistically accessible is that from
the merger of supermassive black holes.  If the SMBH binaries
are all circular, then the expected background has $\Omega_{\rm
gw} \propto f^{2/3}$, in which case the contributions of
longer-wavelength gravitational waves to the observables are
suppressed.  The suppression is even stronger if the SMBH orbits
are eccentric (e.g., the scaling may be as strong as
$\Omega_{\rm gw} \propto f^{3}$) \cite{Enoki:2006kj,Sesana:2008mz,AmaroSeoane:2009af,Kelley:2017lek}.  We thus conclude that the
distant-source limit is valid for the SMBH signal.

\subsection{Chirality}

Here we have taken the normal modes of the transverse-traceless
tensor field to be $\Psi^{k,TE}_{(\ell m)ab}$ and
$\Psi^{k,TB}_{(\ell m)ab}$.  Statistical homogeneity then requires
that these have equal power \cite{Kamionkowski:2014faa}.

However, we could have equally well worked alternatively with a
helicity basis, in terms of TAM modes $\Psi^{k,\pm}_{(\ell m)ab}=
2^{-1/2} \left[ \Psi^{k,TE}_{(\ell m)ab} \pm i\Psi^{k,TB}_{(\ell m)ab}
\right]$.  These two modes represent right- and left-circularly
polarized gravitational waves.  If parity is unbroken, then the
energy densities in the two circular-polarization states should
be the same.  If so, the cross-correlations
$C_\ell^{zB}=C_\ell^{EB}=0$.  

However, it is conceivable, and perhaps even
to be expected, that the stochastic background observed by
pulsar timing and astrometry may be chiral---i.e., may exhibit a
preponderance of one handedness over the other.  The emission
from SMBH binaries is expected to be circularly polarized, to
some degree (depending on the orientation of the binary relative
to the line of sight).  If the background is dominated by a
small number $N$ of SMBHs (see, e.g.,
Refs.~\cite{Sesana:2008xk,Boyle:2010rt,Rosado:2015epa,Kelley:2017lek}),
then the fractional difference
between the powers in the two helicities should be $\sim
N^{-1/2}$; i.e., not too small.  It is thus not advisable to
assume that these cross-correlations will be zero and can thus
be used to as null tests for systematics.  On the other hand, these
parity-breaking power spectra $C_\ell^{zB}$ and $C_\ell^{EB}$ (or
equivalently, the parity-breaking $z\perp$ and $\parallel\perp$
correlation functions) should be pursued observationally along
with the others, as they may shed light on the nature of the
sources that give rise to the background.  Chirality probes that
can be constructed from time-sequence information
\cite{Kato:2015bye} should also be similarly employed.  We moreover note that
these cross-correlations provide pulsar timing and astrometry
with a capability to test the chirality of the
gravitational-wave background in a frequency regime previously
thought to be inaccessible \cite{Smith:2016jqs}.

\section{Conclusions}
\label{sec:conclusions}

Here we have employed a total-angular-momentum formalism to
describe the angular correlations in pulsar-timing and
astrometry probes of a stochastic gravitational-wave
background.  Results were presented both in terms of angular
power spectra and in terms of angular correlation functions and
for all six polarizations that may arise in alternative-gravity
theories.  Redshift-astrometry cross-correlations were provided
for the first time for all six polarizations.  An Appendix
describes an alternative way to rederive simply the
power-spectrum results from plane waves.  The dependence of the
astrometry signal on the frequency spectrum of the
gravitational-wave background was clarified, and it was
speculated that information on the local three-dimensional
metric perturbation might be inferred by the inclusion of
time-sequence information.  We also emphasized
that the parity-breaking cross-correlations, usually assumed to
be zero, will not necessarily vanish for stochastic backgrounds
generated by supermassive-black-hole binaries.

A natural next step would be to ask whether a detected 
gravitational-wave background exhibits any preferred direction. 
One possible way to search for such asymmetries is with bipolar spherical 
harmonics \cite{Hajian:2003qq,Book:2011na,Mingarelli:2013dsa}, 
which could be used to seek, for example, a dipole asymmetry in 
the strength of the gravitational-wave signal.  It may also be 
worthwhile to consider merging the techniques presented here 
with other novel approaches, such as those involving 
gravitational-wave Stokes parameters \cite{Conneely:2018wis}.  
Elaboration of the details and development of such strategies is left 
for future work.

We hope that the mathematical tools and calculational results we
have presented will be of value in further characterization and
exploration of stochastic backgrounds.

\begin{acknowledgments}
We thank E.~Berti for useful discussions.  This work was
supported at Johns Hopkins in part by NASA Grant No.\ NNX17AK38G,
NSF Grant No.\ 0244990, and the Simons Foundation.
L.D. is supported at the Institute for Advanced Study by NASA through
Einstein Postdoctoral Fellowship Grant No.\ PF5-160135 awarded by
the Chandra X-ray Center, which is operated by the Smithsonian
Astrophysical Observatory for NASA under Contract No.\ NAS8-03060.
\end{acknowledgments}

\appendix

\begin{widetext}
\section{Review of total-angular-momentum waves}
\label{sec:TAM}

\subsection{The standard decomposition}

The most general symmetric tensor field
$h_{ab}(\boldsymbol{x})=h_{(ab)}(\boldsymbol{x}) \equiv
\left[h_{ab}(\boldsymbol{x})+h_{ba}(\boldsymbol{x}) \right]/2$ can be decomposed
into a trace component $h(\boldsymbol{x})$, a longitudinal component
$\xi(\boldsymbol{x})$, two vector components $w_a$ (with
$\nabla^a w_a=0$), and two transverse-traceless tensor
components $h_{ab}^{TT}$ (which satisfy 
$\nabla^a h_{ab}^{TT}=0$ and $h^a{}_a=0$), as
\begin{equation}
     h_{ab} = h
     g_{ab}+\left(\nabla_{a}\nabla_{b}-\frac{1}{3}g_{ab}
     \nabla^{2}\right)\xi+\nabla_{(a}w_{b)}+h_{ab}^{TT}.
\label{eqn:SVTdecomposition}
\end{equation}

The most general rank-two symmetric $3\times3$ tensor can
be expanded as
\begin{equation}
     h_{ab}(\boldsymbol{x}) = \sum_{{\boldsymbol{k}},s}
     \varepsilon^s_{ab}(\hat k) h_s(\boldsymbol{k}) e^{i \boldsymbol{k}\cdot\boldsymbol{x}} +cc,
\end{equation}
in terms of Fourier modes of wave vector $\boldsymbol{k}$ 
and in terms of six polarization states
$\varepsilon^s_{ab}(\boldsymbol{k})$, where $s=\{0,z,x,y,+,\times\}$, for
the trace, longitudinal, two vector, and two transverse-traceless
polarizations, respectively, with amplitudes $h_s(\boldsymbol{k})$
\cite{Jeong:2012df}.  The quantity $cc$ denotes the complex conjugate
of the first term.
The polarization tensors are normalized such that
$\varepsilon^{s\,ab} \varepsilon^{s'}_{ab}=2\delta_{ss'}$.  The
two transverse-traceless polarization states that propagate in
general relativity have $k^a \varepsilon^{+,\times}_{ab}=0$. 
Power spectra $P_h(k)$ for these transverse-traceless
gravitational waves are defined by
\begin{equation}
     \VEV{h_s(\boldsymbol{k}) h_{s'}(\boldsymbol{k}')} = \delta_{ss^{\prime}} (2\pi)^3
     \delta_D(\boldsymbol{k}-\boldsymbol{k}') \frac{P_h(k)}{4},
\label{def:pk_tensor}
\end{equation}
for $s,s'=\{+,\times\}$.

To connect with prior work on pulsar timing and astrometry, we
note that with these conventions, the wave number $k=2\pi f$ in
terms of the gravitational-wave frequency $f$, and
\begin{equation}
     P_h(2\pi f) = \frac{3 H_0^2 \Omega_{\rm gw}(f)}{8\pi^3
     f^5} = \frac{1}{2\pi f^2} S_h(f) = \frac{1}{4\pi f^3} h_c^2(f),
\label{eqn:translation}     
\end{equation}
in terms of the contribution $\Omega_{\rm gw}(f)$ per unit
logarithmic frequency interval to the critical density (and
$H_0$ is the Hubble parameter), the gravitational-wave spectral
density $S_h(f)$ \cite{Maggiore:1999vm}, and $h_c(f)$, the
dimensionless amplitude per logarithmic frequency interval.  To
be precise, the total gravitational-wave energy density, summing
over all frequencies, is $\rho_{\rm gw} = \rho_c \int\,(df/f)\,
\Omega_{\rm gw}(f)$, where $\rho_c = 3\, H_0^2/(8\pi G)$ is the
critical density.

\subsection{Total-angular-momentum waves}

TAM waves \cite{Dai:2012bc} provide an
alternative complete orthonormal set of basis functions for
tensor fields.  Here, the Fourier wave vector $\boldsymbol{k}$ is
replaced by quantum numbers $k\ell m$, where $k$ is a wave number and
$\ell m$ are TAM quantum numbers.  The 5 trace-free polarizations
are replaced by 5 sets of TAM modes, which include L (a trace-free
longitudinal mode), VE and VB (two vector modes), and TE and TB
(the two transverse-traceless modes).  We augment the formalism
of Ref.~\cite{Dai:2012bc} to include a trace degree of freedom.
To facilitate comparison with prior astrometry work, we also
construct from the L mode tensor harmonic and the scalar harmonic a
ST mode and a SL mode to accord with those of
Refs.~\cite{Lee:2008aa,OBeirne:2018slh,Mihaylov:2018uqm}.

Then, any symmetric $h_{ab}(\boldsymbol{x})$ can be expanded,
\begin{equation}
     h_{ab}(\boldsymbol{x}) = \sum_{\alpha k\ell m}
     4\pi i^\ell h^{\alpha}_{k\ell m}
     \Psi^{k,\alpha}_{(\ell m)ab}(\boldsymbol{x}) + cc,
\label{eqn:TAMexpansion}
\end{equation}
in terms of TAM waves $\Psi^{k,\alpha}_{(\ell m)ab}(\boldsymbol{x})$.  Here,
$\sum_k$ is a shorthand for $\int k^2\, dk/(2\pi)^3$, and 
$\alpha$ is summed over ST, SL, VE, VB, TE, and TB.  The
VE/VB/TE/TB TAM waves are as given in Ref.~\cite{Dai:2012bc}.

The SL and ST modes are
\begin{align}
    \Psi^{ST}_{(\ell m)ab} (\boldsymbol{x}) &= \frac{\sqrt{2}}{3}
    \Psi_{(\ell m)} (\boldsymbol{x}) g_{ab} + \sqrt{\frac{1}{3}}
    \Psi^{L}_{(\ell m)ab} (\boldsymbol{x}), \\
    \Psi^{SL}_{(\ell m)ab} (\boldsymbol{x}) &= \frac{1}{3}
    \Psi_{(\ell m)} (\boldsymbol{x}) g_{ab} - \sqrt{\frac{2}{3}}
    \Psi^{L}_{(\ell m)ab} (\boldsymbol{x}), 
\end{align}
where $\Psi_{(\ell m)}(\boldsymbol{x}) = j_\ell(kr)Y_{\ell m}(\boldsymbol{\hat n})$ is
the scalar TAM wave and $j_\ell(kr)$ is the spherical Bessel function.
The ST TAM wave $\Psi^{ST}_{(\ell m)ab}
(\boldsymbol{x})$ has components only transverse to the
direction of its gradients, and $\Psi^{SL}_{(\ell m)ab}
(\boldsymbol{x})$ is entirely aligned with the gradient.  These
TAM waves are normalized in accord with the conventions of
Ref.~\cite{Dai:2012bc}.

If the metric
perturbation is constructed of wavelike solutions that propagate
at the speed of light, then the time-dependent metric
perturbation $h_{ab}(\boldsymbol{x},t)$ is obtained by multiplying the
summand in Eq.~(\ref{eqn:TAMexpansion}) by $e^{-ikt}$.  If we are
dealing with a stochastic background of general-relativistic
gravitational waves, then the sum is only over TE/TB modes.  In
a statistically isotropic stochastic background of GR
gravitational waves, the TAM-wave coefficients are statistically
independent and taken from a random distribution with variance
$P_h(k)$; i.e.,
\begin{equation}
     \VEV{ \left(h^\alpha_{k\ell m}\right)^* h^\beta_{k'\ell'm'}}
     = \frac{(2\pi)^3}{2k^2} P_h(k)
     \delta_{\ell\ell'} \delta_{mm'} \delta_{\alpha\beta} \delta(k-k')
\end{equation}
for the TE/TB modes.  In
alternative-gravity theories, there will be analogous
expressions for VE/VB, SL, and ST modes in terms of vector and
scalar power spectra, if such modes exist and propagate.
Note that the TE/TB modes exist only for $\ell\geq2$ and the VE/VB
modes for $\ell\geq1$.  Note also that statistical homogeneity
requires that $P_{TE}(k)=P_{TB}(k)$ and $P_{VE}(k)=P_{VB}(k)$
\cite{Kamionkowski:2014faa}, but the power spectra for SL and ST
may most generally differ.

Reference~\cite{Dai:2012bc} provides an array of results on the
properties of these TAM waves, related scalar and vector TAM
waves, and several alternative TAM-wave bases.  In particular,
Eq.~(94) in that paper provides the projections of these TAM
waves onto an orthonormal basis determined by unit vectors in
the radial ($\boldsymbol{\hat{n}}$) and angular
($\boldsymbol{\hat{\theta}}$, and $\boldsymbol{\hat{\phi}}$) directions 
in the usual spherical coordinates.  The central quantities we will
need for this work are $n^b  \Psi^{\alpha,k}_{(\ell m)ab}(\boldsymbol{x})$.
From Eq.~(94) of Ref.~\cite{Dai:2012bc}, and our definition of
the SL and ST modes, we have
\begin{eqnarray}
    \hat{n}^a \Psi^{SL}_{(\ell m)ab} (\boldsymbol{x})
		&=& \sqrt{\ell (\ell+1)} \frac{1}{kr} \left(
                j'_\ell(kr) - \frac{j_\ell(kr)}{kr} \right)
                Y^E_{(\ell m)b} (\boldsymbol{\hat{n}})  - \left[ 2
                \frac{j'_\ell(kr)}{kr} + \left( 1-
                \frac{\ell (\ell+1)}{(kr)^2} \right) j_\ell (kr) \right]   
                Y^L_{(\ell m)b} (\boldsymbol{\hat{n}}),
                \nonumber \\
     \hat{n}^a \Psi^{ST}_{(\ell m)ab} (\boldsymbol{x})
		&=& - \frac{1}{\sqrt{2}} \left[ \sqrt{\ell (\ell+1)}
                \frac{1}{kr} \left( 
                j'_\ell(kr) - \frac{j_\ell(kr)}{kr} \right)
                Y^E_{(\ell m)b} (\boldsymbol{\hat{n}}) - \left( 2
                \frac{j'_\ell(kr)}{kr}
                -  
                \frac{\ell (\ell+1)}{(kr)^2} j_\ell(kr) \right)
                Y^L_{(\ell m)b} (\boldsymbol{\hat{n}}) \right],
                \nonumber \\
      \hat{n}^a \Psi^{VE}_{(\ell m)ab} (\boldsymbol{x}) &=&
     \sqrt{2}\left[ \frac{j_\ell'(kr)}{kr}  + \left(
     \frac{1}{2} + \frac{(1-\ell-\ell^2)}{(kr)^2}
     \right) j_\ell(kr) \right] Y^E_{(\ell m)b} (\boldsymbol{\hat{n}})
      - \frac{\sqrt{2\ell (\ell+1)}}{kr} \left(
     j'_\ell(kr) - \frac{j_\ell(kr)}{kr} \right) Y^L_{(\ell m)b}
     (\boldsymbol{\hat{n}}),\nonumber \\
     \hat{n}^a \Psi^{VB}_{(\ell m)ab} (\boldsymbol{x}) &=&
	- \frac{i}{\sqrt{2}} \left( j'_{\ell}(kr) - \frac{j_{\ell}
        (kr)}{kr} \right) Y^B_{(\ell m)b}(\boldsymbol{\hat{n}}),
        \nonumber \\
     \hat{n}^a \Psi^{TE}_{(\ell m)ab} (\boldsymbol{x}) &=& -
       N_\ell  \frac{j_\ell(kr)}{(kr)^2}
       Y^L_{(\ell m)b}(\boldsymbol{\hat{n}}) - \frac{
       N_\ell}{\sqrt{\ell (\ell+1)}} \left( \frac{j_\ell'(kr)}{kr} +
       \frac{j_\ell(kr)}{(kr)^2} \right)
       Y^E_{(\ell m)b}(\boldsymbol{\hat{n}}), \nonumber \\
     \hat{n}^a \Psi^{TB}_{(\ell m)ab} (\boldsymbol{x}) &=&
       -i\frac{N_\ell}{\sqrt{\ell (\ell+1)}} \frac{j_\ell(kr)}{kr}
       Y^B_{(\ell m)b}(\boldsymbol{\hat{n}}),
\label{eqn:centralresult}       
\end{eqnarray}
with $N_\ell = \sqrt{(\ell+2)!/[2(\ell-2)!]}$.
Note that these vectorial quantities have a projection
onto a vector spherical harmonic $Y^L_{(\ell m)a}(\boldsymbol{\hat{n}}) \equiv -n_a
Y_{\ell m}(\boldsymbol{\hat{n}})$, which points along the radial direction, and
another onto either the vector spherical harmonic
$Y^E_{(\ell m)a}(\boldsymbol{\hat{n}})$ or $Y^B_{(\ell
m)a}(\boldsymbol{\hat{n}})$, which lie in the plane of the sky.
We thus define radial functions $R_\ell^{L,\alpha}(x)$ and
$R_\ell^{E,\alpha}(x)$ through
\begin{equation}
    \hat{n}^a \Psi^{\alpha}_{(\ell m)ab} (\boldsymbol{x}) =
       R_\ell^{L,\alpha}(kr)
       Y^L_{(\ell m)b}(\boldsymbol{\hat{n}}) + R_\ell^{E,\alpha}(kr)
       Y^E_{(\ell m)a}(\boldsymbol{\hat{n}}),
\end{equation}
and analogously for $R_\ell^{B,\alpha}(x)$.

We will also need in our calculations the
quantities  $n^a n^b  \Psi^{\alpha,k}_{(\ell m)ab}(\boldsymbol{x})$, which
are obtained from the above expressions by replacing
$Y^L_{(\ell m)a}(\boldsymbol{\hat{n}})$ by $-Y_{\ell m}(\boldsymbol{\hat{n}})$ and ignoring the E/B
components, given the orthogonality of the E/B vector spherical
harmonics to the radial direction $\boldsymbol{\hat{n}}$.

\section{Useful Legendre-polynomial relations}
\label{sec:legendretricks}

Here we show how to derive the Hellings-Downs curve from a power
spectrum $C_\ell \propto (\ell-2)!/(\ell+2)!$ for $\ell\geq2$ and from the
fact that $\textrm{HD}(\Theta) = \sum_\ell (2\ell+1) C_\ell
P_\ell(\cos\Theta)$.  Expanding this quantity using partial
fractions yields
\begin{equation}
     \textrm{HD}(\Theta) = \sum_{\ell=2}^{\infty} (2\ell+1)
     \frac{(\ell-2)!}{(\ell+2)!} P_\ell(\cos\Theta) =  \frac{1}{2}
     \sum_{\ell=2}^{\infty} \left( \frac{1}{\ell-1} - \frac{1}{\ell} -
     \frac{1}{\ell +1} + \frac{1}{\ell+2} \right) P_\ell(\cos\Theta).
\end{equation}
Each of these four infinite sums can be calculated using the
generating function of the Legendre polynomials, which is given
by
\begin{equation}
     \frac{1}{\sqrt{t^2 - 2tx + 1}} = \sum_{n=0}^{\infty} t^n P_n(x).
\end{equation}
For example, let us rewrite the first partial fraction as
\begin{equation}
     \frac{1}{\ell-1} = \int_{0}^{\infty} e^{-z(\ell-1)} dz.
\end{equation}
Then we can rewrite the sum as
\begin{align}
     \sum_{\ell=2}^{\infty} \frac{1}{\ell-1} P_\ell(\cos\Theta) &=
     \int_{0}^{\infty} e^z \sum_{\ell=2}^{\infty} e^{-z\ell}
     P_\ell(\cos\Theta) \, dz = \int_{0}^{\infty} e^z \left(\frac{1}{\sqrt{e^{-2z} -
     2xe^{-z} + 1}} -1 -e^{-z} x \right) \, dz \nonumber \\
     &= \int_{1}^{\infty} \left(\frac{y}{\sqrt{1 - 2xy + y^2}}
     -1 - \frac{x}{y} \right) \, dy = 1 - x - \sqrt{2-2x} + x
     \ln \left( \frac{2}{1-x + \sqrt{2-2x}} \right),
\end{align}
where $x=\cos\Theta$.
Similar calculations for the other sums give
\begin{align}
     \sum_{\ell=2}^{\infty} \frac{1}{\ell} P_\ell(\cos\Theta) &= - x +
     \ln \left( \frac{2}{1-x + \sqrt{2-2x}} \right) \\
     \sum_{\ell=2}^{\infty} \frac{1}{\ell+1} P_\ell(\cos\Theta) &= -1 -
     \frac{1}{2} x + \ln \left( 1+ \sqrt{\frac{2}{1-x}} \right)
     \\
     \sum_{\ell=2}^{\infty} \frac{1}{\ell+2} P_\ell(\cos\Theta) &= -
     \frac{3}{2} - \frac{1}{3} x + \sqrt{2-2x} + x \ln \left( 1+
     \sqrt{\frac{2}{1-x}} \right).
\end{align}
We then obtain the Hellings-Downs angular correlation function,
\begin{equation}
     \textrm{HD}(\Theta) = \frac{1}{4} + \frac{1}{12} x + \frac{1}{2} \ln
     \left( \frac{1-x}{2} \right) - \frac{x}{2} \ln \left(
     \frac{1-x}{2} \right)
     = \frac{1}{2} (1-x) \log \left[\frac{1}{2}
     (1-x)\right] - \frac{1}{6} \left[\frac{1}{2}
     (1-x)\right] + \frac{1}{3}.
\label{eqn:HD}     
\end{equation}

\section{Alternative derivation of power spectra}
\label{sec:HDsimple}

Here we present an alternative derivation of the redshift
and angular-deflection power spectra (see also
Ref.~\cite{Roebber:2016jzl}).  The calculation
begins with the well-known angular dependence,\footnote{Note
that this function has an unphysical discontinuity at
$\cos\theta\to -1$.  This is smoothed by the source term.  It
can be shown that the neglect of the source term has no effect
on the subsequent derivation, though.}
\begin{equation}
     z(\boldsymbol{\hat{n}}) = \frac{n^a n^b
     h_{ab}}{2(1+\boldsymbol{\hat{p}}\cdot \boldsymbol{\hat{n}})}
\end{equation}
of the redshift in the presence of a gravitational wave
traveling in the $\boldsymbol{\hat{p}}$ direction.  For example, for a
transverse-traceless gravitational wave in the
$\boldsymbol{\hat{z}}$ direction with $+$ polarization, this
becomes
\begin{equation}
     z(\boldsymbol{\hat{n}}) \propto (1-\cos\theta) \cos2\phi.
\end{equation}     
The spherical-harmonic coefficients for this angular pattern are
\begin{equation}
     z_{\ell m} = \int \, d\boldsymbol{\hat{n}}\, Y_{\ell
     m}(\boldsymbol{\hat{n}}) z(\boldsymbol{\hat{n}}) \propto
     \sqrt{\frac{(2\ell+1)(\ell-2)!}{(\ell+2)!}} \left(
     \delta_{m2} + \delta_{m,-2} \right).
\end{equation}
The contribution of this mode to the power spectrum is thus
$\propto \sum_m |z_{\ell m}|^2/(2\ell+1) \propto (\ell-2)!/(\ell+2)!$.  Since
this is a rotational invariant, the contribution of any Fourier
mode in any direction (and of any magnitude), and of either
polarization, is the same.  From this we infer that $C_\ell \propto
(\ell-2)!/(\ell+2)!$.  The power spectra for the vector and
scalar modes can be similarly obtained.

Likewise, the angular deflection from a wave propagating in the
$\boldsymbol{\hat{p}}$ direction is \cite{Book:2010pf}
\begin{equation}
     (\delta n)^a (\boldsymbol{\hat{n}}) = \frac{(n^a+p^a)
     n^b n^c h_{bc}}{2(1+\boldsymbol{\hat{n}} \cdot
     \boldsymbol{\hat{p}})} - \frac{1}{2} n^b h_{ab}.
\end{equation}
The scalar E-mode pattern associated with this is
$E(\boldsymbol{\hat{n}}) = \nabla_a (\delta n)^a$, while the
B-mode pattern is $B(\boldsymbol{\hat{n}})=
\epsilon_{abc} n_a \nabla_b (\delta n)^c$.  Using $\nabla_a n_b
= \delta_{ab}- n_a n_b$ and $\nabla_a(\boldsymbol{\hat{p}} \cdot
\boldsymbol{\hat{n}}) = p_a - n_a(\boldsymbol{\hat{p}} \cdot
\boldsymbol{\hat{n}})$, we find
\begin{equation}
     E(\hat n) = -\frac{1}{2} \textrm{Tr\,}h + \frac{(n^a+p^a) n^b
     h_{ab}}{1 + \hat p \cdot \hat n},
\label{eqn:E}     
\end{equation}
and
\begin{equation}
      B(\boldsymbol{\hat{n}}) = \epsilon_{abc}
      \frac{p^a n^d n_c h_{bd} }{1+\hat p \cdot \hat n}.
\label{eqn:B}      
\end{equation}

Now consider again the transverse-traceless gravitational wave
propagating in the $\boldsymbol{\hat{z}}$ direction with $+$
polarization.  The transverse-traceless wave has
$\textrm{Tr}\,h=0$ and $p^a h_{ab}=0$ from which we infer that
{\it the angular pattern of the E mode from transverse-traceless
gravitational waves is identical with that for the redshift}.
This thus explains why the E-mode correlation function
$\beta^{EE}(\Theta)$ has the exact same form as the Hellings-Downs
curve.  It is furthermore found that the B-mode pattern is the
same as the E-mode pattern, but rotated about
$\boldsymbol{\hat{p}}$ by $45^\circ$, thus explaining why the
B-mode correlation function and power spectrum are the same as
those for the E mode (and also why it is not correlated with the
redshift).  The power spectra for the vector and scalar E and B
modes are similarly derived.

\section{Relation between deflection-angle correlation functions
     and power spectra}
\label{sec:angularcorrelations}

The correlation functions described in Sec.
\ref{sec:correlations} are rotationally invariant.  We can
evaluate them most easily, though, by choosing one of the two
points to be correlated to be at the north pole
($\Theta=\Phi=0$) and the other at a $(\Theta,\Phi=0)$.
In terms of the conventional scalar spherical harmonics,
$Y_{\ell m}(\Theta,\Phi)$,
the vector spherical harmonics are~\cite{Dai:2012bc}
\begin{align}
  Y^{E}_{(\ell m)a}(\Theta,\Phi)
  =& -\frac{r}{\sqrt{\ell (\ell+1)}} \nabla_a Y_{\ell m}(\Theta,\Phi)
  = -\frac{1}{\sqrt{\ell (\ell+1)}} \left[
    \hat{\theta}_a \frac{\partial}{\partial\Theta}
    + \hat{\phi}_a \frac{1}{\sin\Theta} \frac{\partial}{\partial\Phi}
    \right] Y_{\ell m}(\Theta,\Phi) \nonumber\\
  =& -\frac{1}{2} \frac{1}{\sqrt{\ell (\ell+1)}} \left\{
  \hat{\theta}_a \left[ \sqrt{(\ell-m)(\ell+m+1)} e^{-i\Phi}
  Y_{\ell,m+1}(\Theta,\Phi)
    - \sqrt{(\ell+m)(\ell-m+1)} e^{i\Phi}
    Y_{\ell,m-1}(\Theta,\Phi) \right] \right. \nonumber \\
  &  \left. + \hat{\phi}_a \frac{2im}{\sin\Theta} Y_{\ell,m}(\cos\Theta)
  \right\}, \\
  Y^{B}_{(\ell m)a}(\Theta,\Phi)
  =& -\frac{r}{\sqrt{\ell (\ell+1)}} \epsilon_{abc} n^b \nabla^c Y_{\ell m}(\Theta,\Phi)
  = -\frac{1}{\sqrt{\ell (\ell+1)}} \left[
    \hat{\phi}_a \frac{\partial}{\partial\Theta}
    - \hat{\theta}_a \frac{1}{\sin\Theta} \frac{\partial}{\partial\Phi}
    \right] Y_{\ell m}(\Theta,\Phi) \nonumber\\
  =& -\frac{1}{2} \frac{1}{\sqrt{\ell (\ell+1)}} \left\{
  \hat{\phi}_a \left[ \sqrt{(\ell-m)(\ell+m+1)} e^{-i\Phi} Y_{\ell,m+1}(\Theta,\Phi)
    - \sqrt{(\ell+m)(\ell-m+1)} e^{i\Phi}
    Y_{\ell,m-1}(\Theta,\Phi) \right] \right. \nonumber \\
  &  \left. - \hat{\theta}_a \frac{2im}{\sin\Theta} Y_{\ell,m}(\cos\Theta)
  \right\} .
\end{align}
There is a third vector spherical harmonic $Y^{L}_{(\ell m)a}(\Theta,\Phi)$
in the direction normal to the two-sphere of the sky, but it
does not enter our calculations here.
These vector spherical harmonics obey the orthogonality relation
\begin{equation}
  \int d\mathbf{\hat{n}}\, Y^{X}_{(\ell m) a}(\mathbf{\hat{n}})
  \left[Y^{X' a}_{(\ell^\prime m')}\right]^\ast(\mathbf{\hat{n}})
  = \delta_{\ell\ell^\prime} \delta_{mm'} \delta_{XX'},
  \label{eqn:vectorY-orthogonality}
\end{equation}
where $X,X' = \{E,B,L\}$.

Evaluating the vector spherical harmonics at $\Phi=0$ gives
\begin{align}
  Y^{E}_{(\ell m)a}(\Theta,\Phi=0)
  =& -\frac{1}{2} \sqrt{\frac{1}{\ell
  (\ell+1)}\frac{2\ell+1}{4\pi}\frac{(\ell-m)!}{(\ell+m)!}}
  \left\{ \hat{\theta}_a \left[
    P_\ell^{m+1}(\cos\Theta) -
    (\ell+m)(\ell-m+1)P_\ell^{m-1}(\cos\Theta)\right] \right. \nonumber\\
  & \left. + \hat{\phi}_a \frac{2im}{\sin\Theta} P_\ell^m(\cos\Theta)
  \right\}, \\
  Y^{B}_{(\ell m)a}(\Theta,\Phi=0)
  =& -\frac{1}{2} \sqrt{\frac{1}{\ell
  (\ell+1)}\frac{2\ell+1}{4\pi}\frac{(\ell -m)!}{(\ell+m)!}}
  \left\{ \hat{\phi}_a
  \left[ P_\ell^{m+1}(\cos\Theta) -
  (\ell+m)(\ell-m+1)P_\ell^{m-1}(\cos\Theta)\right] \right. \nonumber \\
  & \left. - \hat{\theta}_a \frac{2im}{\sin\Theta} P_\ell^m(\cos\Theta)
  \right\} ,
\end{align}
where we have expressed the scalar spherical harmonics,
\begin{equation}
  Y_{\ell m} (\Theta,\Phi) = \sqrt{\frac{2\ell+1}{4\pi}
  \frac{(\ell-m)!}{(\ell +m)!}} P^m_\ell (\cos\Theta) e^{im\Phi} ,
\end{equation}
in terms of associated Legendre polynomials, $P^m_\ell (\cos\Theta)$.
Further evaluating at $\Theta=0$
using Eq.~(5.2) in Ref.~\cite{Kamionkowski:1996ks} gives
\begin{align}
  Y^E_{(\ell m)a} (0,0)
  &= \frac{1}{2} \sqrt{\frac{2\ell+1}{4\pi}} \left[(\delta_{m1} -
    \delta_{m,-1}) \hat{\theta}_a + i (\delta_{m1} +
    \delta_{m,-1}) \hat{\phi}_a \right], \\
  Y^B_{(\ell m)a} (0,0) &= \frac{1}{2} \sqrt{\frac{2\ell+1}{4\pi}} \left[
    (\delta_{m1} - \delta_{m,-1}) \hat{\phi}_a
    -i(\delta_{m1} + \delta_{m,-1}) \hat{\theta}_a \right].
\end{align}
With these expressions, the correlation functions become,
\begin{align}
  C^{\parallel\parallel} (\Theta)
  &= \sum_{\ell m} \left[
    C_\ell^{EE}  Y_{(\ell m)\theta}^{E}(\Theta,0) Y_{(\ell m)\theta}^{E*}(0,0)
    + C_\ell^{BB}  Y_{(\ell m)\theta}^{B}(\Theta,0) Y_{(\ell m)\theta}^{B*}(0,0)
    \right] \nonumber\\
  &= \frac{1}{2} \sum_{\ell} \frac{2\ell+1}{4\pi} \left\{
  C_\ell^{EE} \left[P_\ell^0(\cos\Theta) - \frac{1}{\ell (\ell+1)} P_\ell^2(\cos\Theta)\right]
    - C_\ell^{BB} \frac{2}{\ell (\ell+1)} \frac{1}{\sin\Theta}
    P_\ell^1(\cos\Theta) \right\} \\
  C^{\perp\perp} (\Theta)
  &= \sum_{\ell m} \left[
    C_\ell^{EE}  Y_{(\ell m)\phi}^{E}(\Theta,0) Y_{(\ell m)\phi}^{E*}(0,0)
    + C_\ell^{BB}  Y_{(\ell m)\phi}^{B}(\Theta,0) Y_{(\ell m)\phi}^{B*}(0,0)
    \right] \nonumber\\
  &= \frac{1}{2} \sum_{\ell} \frac{2\ell+1}{4\pi} \left\{
  - C_\ell^{EE} \frac{2}{\ell (\ell+1)} \frac{1}{\sin\Theta} P_\ell^1(\cos\Theta)
  + C_\ell^{BB} \left[P_\ell^0(\cos\Theta) - \frac{1}{\ell (\ell+1)} P_\ell^2(\cos\Theta)\right]
  \right\},
\end{align}
while the cross-correlation function becomes
\begin{align}
  C^{\perp\parallel}(\Theta) &= \sum_{\ell m} \left[
    C_\ell^{EE} Y_{(\ell m)\phi}^{E}(\Theta,0)  Y_{(\ell m)\theta}^{E*}(0,0)
    + C_\ell^{BB} Y_{(\ell m)\phi}^{B}(\Theta,0)  Y_{(\ell m)\theta}^{B*}(0,0)
    \right] \nonumber\\
  &= -\frac{1}{2} \sum_{\ell} \frac{2\ell+1}{4\pi} C_\ell^{EB}
  \left[\frac{1}{\ell(\ell+1)} P_\ell^2(\cos\Theta) - P_\ell^0(\cos\Theta)
    + \frac{2}{\ell(\ell+1)}\frac{1}{\sin\Theta} P_\ell^1(\cos\Theta) \right].
\end{align}
The $z\parallel$ and $z\perp$ correlations are analogously
derived most simply by putting the deflection at the north pole
and the redshift at $(\Theta,0)$.

To write the power spectra in terms of the correlation
functions, we define the vector quantities
\begin{align}
  \mathcal{C}_a(\Theta) &\equiv C^{\parallel\parallel} (\Theta) \hat{\theta}_a
  +i C^{\perp\perp} (\Theta) \hat{\phi}_a = \sum_{\ell}
  \sqrt{\frac{2\ell+1}{4\pi}} \left[ C^{EE}_\ell Y^{E}_{(\ell1) a} (\Theta,0)
    + i C^{BB}_\ell Y^{B}_{(\ell1) a} (\Theta,0) \right] \\
  \mathcal{D}_a(\Theta) &\equiv C^{\perp\parallel}(\Theta)
  \left(\hat{\theta}_a - i\hat{\phi}_a\right)
  = \sum_{\ell} \sqrt{\frac{2\ell+1}{4\pi}} C_\ell^{EB}
  \left[Y_{(\ell 1)a}^{E}(\Theta,0) - iY_{(\ell 1)a}^{B}(\Theta,0)\right].
\end{align}
Using the orthogonality relation in Eq.~\eqref{eqn:vectorY-orthogonality},
which holds if the vector spherical harmonics are evaluated at $\Phi=0$,
the multipole moments are
\begin{equation}
     C^{EE}_\ell = \sqrt{\frac{4\pi}{2\ell+1}} \int d \boldsymbol{\hat{n}} \,
     \mathcal{C}_a (\Theta) Y^{Ea}_{(\ell1)} (\Theta,0) \qquad
     C^{BB}_\ell = -i \sqrt{\frac{4\pi}{2\ell+1}} \int
        d \boldsymbol{\hat{n}} \, \mathcal{C}_a (\Theta) Y^{Ba}_{(\ell1)}
        (\Theta,0),
\label{eqn:EBinverses}
\end{equation}
and from these follow Eqs.~(\ref{eqn:inverserelationEE}) and
(\ref{eqn:inverserelationBB}). Similarly, for the cross-correlation,
\begin{equation}
  C_\ell^{EB} = \frac{1}{2} \sqrt{\frac{4\pi}{2\ell+1}} \int
  d \boldsymbol{\hat{n}} \, \mathcal{D}_a(\Theta)
  \left[Y_{(\ell 1)}^{Ea}(\Theta,0) + iY_{(\ell 1)}^{Ba}(\Theta,0)\right] ,
\end{equation}
from which Eq.~\eqref{eqn:inverserelationEB} follows.
The inverse relations in
Eq.~(\ref{eqn:crossinverse}) for the redshift-deflection
cross-correlations follow from the orthogonality of the
spherical harmonics.
\end{widetext}

\end{document}